\documentclass[aps,pre,superscriptaddress,longbibliography,reprint,11pt,tightenlines]{revtex4-2}

\usepackage[english]{babel}

\usepackage{graphicx}
\usepackage{bm}
\usepackage{amsmath}
\usepackage{amssymb}
\usepackage{amsfonts}
\usepackage{xcolor}
\usepackage{natbib}

\usepackage{xr}
\usepackage{xr-hyper}
\usepackage{hyperref} 
\hypersetup{nolinks=true} 

\newcommand*\bsa{\boldsymbol{\alpha}}
\newcommand*\bsb{\boldsymbol{\beta}}

\begin{document}
\title{\Large Odd electrical circuits}

\author{Harry Walden}
\affiliation{Department of Mathematics, Massachusetts Institute of Technology, Cambridge, MA, USA}

\author{Alexander Stegmaier}
\affiliation{Institut f\"ur Theoretische Physik und Astrophysik, Universit\"at W\"urzburg, W\"urzburg, Germany}

\author{J\"orn Dunkel}
\affiliation{Department of Mathematics, Massachusetts Institute of Technology, Cambridge, MA, USA}

\author{Alexander Mietke}
\email{alexander.mietke@physics.ox.ac.uk}
\affiliation{Department of Physics, University of Oxford, Oxford, UK}

\date{\today}

\begin{abstract}
Non-reciprocal interactions in elastic media give rise to rich non-equilibrium behaviors, but controllable experimental realizations of such odd elastic phenomena remain scarce. Building on recent breakthroughs in electrical analogs of non-Hermitian solid-state systems, we design and analyze scalable odd electrical circuits (OECs) as exact analogs of an odd solid. We show that electrical work can be extracted from OECs via cyclic excitations and trace the apparent energy gain back to active circuit elements. We show that OECs host oscillatory modes that resemble recent experimental observations in living chiral crystals and identify active resonances that reveal a perspective on odd elasticity as a mechanism for mechanical amplification.
\end{abstract}
\maketitle

\section{Introduction}

The extension of classic linear elasticity by non-reciprocal ``odd" stress-strain couplings~\cite{Scheibner:2020gm,fruch23rev,coul17} gives rise to active metamaterials that can exhibit chiral waves~\cite{Scheibner:2020gm}, non-Hermitian edge modes~\cite{scheib20,Fossati2024} and topological defect motility~\cite{braverman2020topological,ponc22}. Non-reciprocal microscopic interactions~\cite{true63} should also enable an extraction of work from odd materials via cyclic excitations, which has inspired ideas of optimal odd engines~\cite{etien2021b,Sousl2021} and may play a role in catalytic micromachines~\cite{koba23}, but a quantitative experimental demonstration remains~\hbox{challenging}.

Theoretical proposals to construct odd solids and study this rich non-equilibrium physics include rotor-spring constructions~\cite{Scheibner:2020gm} and collections of active piezoelectric elements \cite{cheng21}. Robotic metamaterials~\cite{Brandenbourger2019,Chen2021,brandenbourger2022}, colloidal systems~\cite{bili21,tan2022} and filaments \cite{Ishimoto2023,Shankar2024,Surowka2023}, both synthetic and biological, have been shown to exhibit features of odd materials. However, the need to precisely control microscopic energy injection in a system of many interacting agents is currently a major bottleneck for the design of easily tunable and scalable realizations of odd elastic systems.

Here, we describe the construction of odd electrical circuits (OECs), a robust, scalable analog of an odd elastic material. Mechanical-electrical analogies were first used by Maxwell to explain novel electrical phenomena~\cite{olson44}, with later authors completing the correspondence~\cite{fire33}. We build on recent work in topolectrical and non-Hermitian circuits~\cite{kotwal21,steg21} to design a system that encodes physical interactions in exact analogy to a linear odd elastic spring network, bypassing the need for robotic microcontrollers and benefiting from the ability to set microscopic system parameters with high precision. Circuit simulations with realistic losses and tolerances verify the emergence of hallmark odd properties in quantitative agreement with mean-field predictions. Performing the electrical analog of an oscillatory piston compression with OECs reveals characteristic amplification properties and chiral excitations resembling those recently seen in experiments~\cite{chao2024}.

\begin{figure*}
\includegraphics[width=\textwidth]{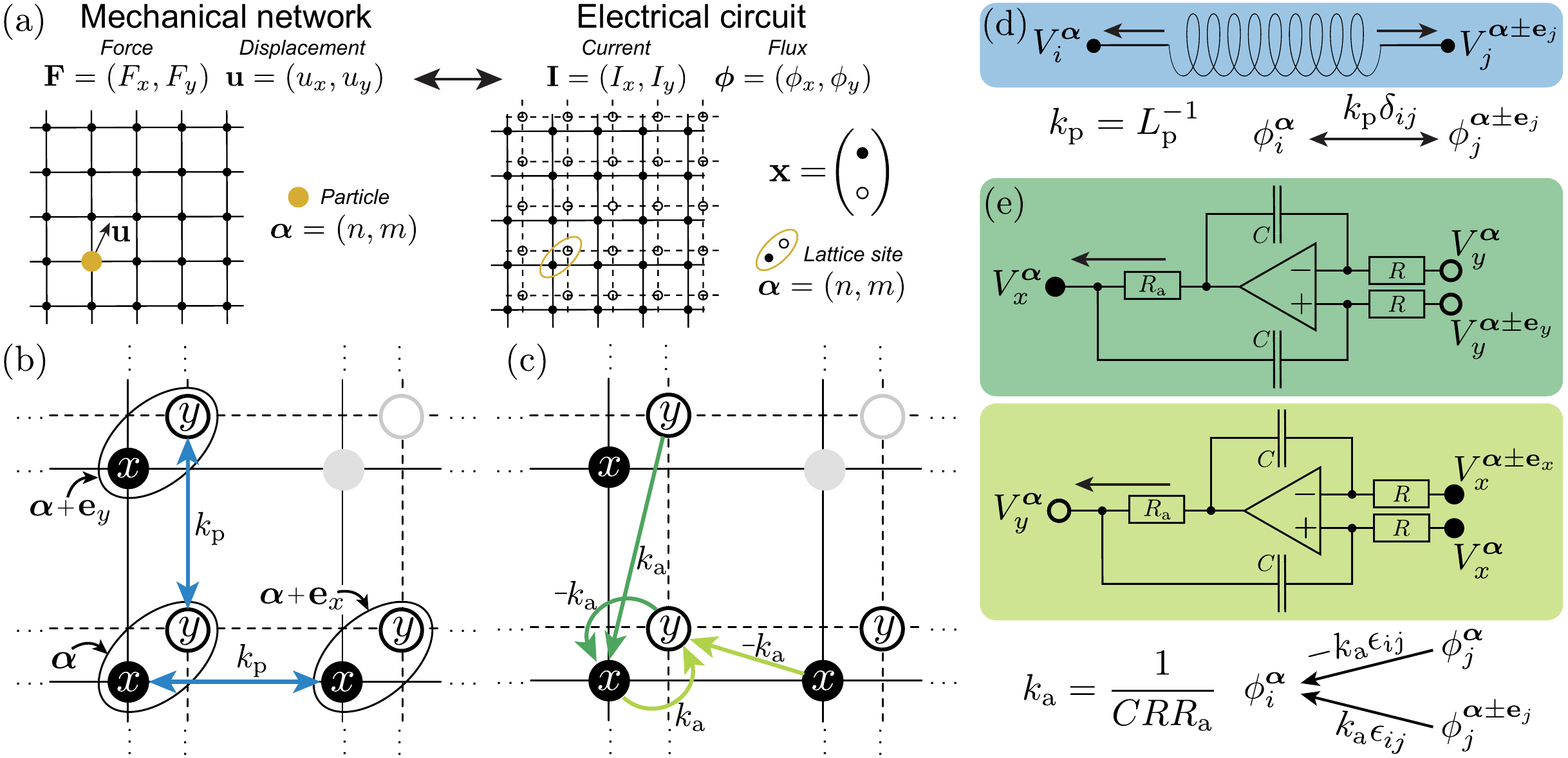}
\caption{Electronic implementation of odd springs. (a)~Vector quantities at lattice site~$\bsa$ correspond in the electrical domain to scalar quantities at a pair of nodes $(\bsa,i )_{i=x,y}$ following the mobility analogy~[see Eq.~(\ref{eq:analog})]. (b)~Classical spring interactions are replaced by two-way inductors with inductance~\hbox{$L_{\text{p}}=k_{\text{p}}^{-1}$} [see Eq.~(\ref{eq: lambda passive})]. (c)~Analog odd spring interactions require non-mutual current flows and effectively negative one-way inductances [see Eqs.~(\ref{eq: lambda active 1}), (\ref{eq: lambda active 2})]. Arrowheads in (b) and~(c) indicate nodes through which a current flows if a flux difference is present. (d),(e)~Realizations of classical (d) and odd (e) spring-analog interactions as electronic circuit elements. Operational amplifiers (OAs) act as ``microscopic'' energy sources to facilitate odd electric interactions.}
\label{Fig:Circuit}
\end{figure*}

\section{Results}
The construction of OECs is based on two-dimensional odd elastic springs that respond to deformations with a force~\cite{Scheibner:2020gm}
\begin{equation}
\mathbf{F}(\mathbf{r})=-(r-r_0)(k_{\text{p}}\hat{\mathbf{r}}+k_{\text{a}}\hat{\mathbf{r}}^{\perp}),\label{eq:OddSpr}
\end{equation}
where $r_0$ is the rest length, $k_{\text{p}}$ is a passive stiffness, \hbox{$\hat{\mathbf{r}}^{\perp}=\boldsymbol{\epsilon}\cdot\hat{\mathbf{r}}$} and $\boldsymbol{\epsilon}$ is the two-dimensional Levi-Civita tensor. The transverse response, proportional to the active stiffness $k_{\text{a}}$, cannot be derived from a potential and therefore requires an external source of energy. The collective properties of a network of particles interacting according to Eq.~(\ref{eq:OddSpr}) can be described in terms of odd elastic moduli~\cite{Scheibner:2020gm,fruch23} that complement the equilibrium response of conventional elastic solids.

We consider a unit-spaced square lattice of particles with nearest-neighbor interactions and label lattice sites by $\bsa \in \mathbb{Z}^2$. Neglecting geometric nonlinearities, Cartesian components of the force acting on a particle at site $\bsa$ are then given by
\begin{equation}\label{eq:NetwForces}
F_i^{\bsa}=-\hspace{-0.2cm}\sum_{\substack{s\in\{\pm 1\}\\j=x,y}}\hspace{-0.1cm}\left(k_{\text{p}}\delta_{ij}+k_{\text{a}}\epsilon_{ij}\right)(u_j^{\bsa}-u_j^{\bsa+s\mathbf e_j}),
\end{equation}
where $u_j^{\bsa}$ denotes the displacement in the $j$\textsuperscript{th} direction of the particle at lattice site $\bsa$ from equilibrium. 

\subsubsection{Mobility analogy and non-reciprocity}

The most convenient analogy to encode Eq.~(\ref{eq:NetwForces}) in an electrical circuit is the mobility analogy~\cite{fire33},
\begin{equation}\label{eq:analog}
F_i^{\bsa} \leftrightarrow I_i^{\bsa}\hspace{1cm}
u_i^{\bsa} \leftrightarrow \phi_i^{\bsa}=\int V^{\bsa}_idt,
\end{equation}
which identifies forces and displacements with currents $I_i^{\bsa}$ and fluxes $\phi_i^{\bsa}$, respectively, and ensures that (mechanical) work has the correct physical representation in the electrical domain: $dW_{\text{mech}}=F_i^{\bsa}du_i^{\bsa}\Leftrightarrow dW_{\text{el}}=I_i^{\bsa}d\phi_i^{\bsa}=I_i^{\bsa}V_i^{\bsa}dt$. To realize vector quantities that respond equivalently to Eq.~(\ref{eq:NetwForces}) in an OEC, we extend each lattice site $\bsa$ to a pair of nodes $(\bsa,i)_{i=x,y}$~(Fig.~\ref{Fig:Circuit}a) at which incoming currents $I_i^{\bsa}$ are determined by flux differences between neighboring nodes. This identification implies an analogy to Eq.~(\ref{eq:NetwForces}) in terms of scalar variables of the form
\begin{equation}
I_i^{\bsa}=-\hspace{-0.2cm}\sum_{\substack{s\in\{\pm 1\}\\j=x,y}}\hspace{-0.1cm}\left(k_{\text{p}}\delta_{ij}+k_{\text{a}}\epsilon_{ij}\right)(\phi_j^{\bsa}-\phi_j^{\bsa+s\mathbf e_j}).\label{eq:NetwCurrsNew}
\end{equation}

We adopt the terminology introduced in~\cite{braverman2020topological} to distinguish \textit{reciprocity} and \textit{mutuality}. Reciprocity is understood in terms of Maxwell-Betti reciprocity in the mechanical domain~\cite{Scheibner:2020gm,fruch23,Chen2021} or Lorentz reciprocity in the electrical domain~\hbox{\cite{olson44,Ballantine1929}}, which both require the underlying systems to be passive. \textit{Mutuality} describes forces or currents which are equal and opposite among interacting objects, namely mechanical bodies (Newton's third law) or electrical nodes (Kirchhoff's current law). Because of the identification of mechanical and electrical work in the mobility analogy Eq.~(\ref{eq:analog}), the breaking of Maxwell-Betti reciprocity in odd elastic systems implies non-reciprocal effects must also be present in OECs. Consequently, we expect that passive electrical components must be complemented by active circuit elements in order to realize an OEC.

\subsubsection{Design principles}

A circuit response is generally captured by an admittance matrix~\cite{saadat1999power} that must in our case mediate flux-current couplings. We therefore seek now a matrix~\smash{$\Lambda^{\bsa \bsb}_{ij}$} of inverse inductances such that currents arriving at node~$(\bsa,i)$,
\begin{equation}
    I_i^{\bsa} = -\sum_{j,\bsb} \Lambda^{\bsa \bsb}_{ij} (\phi_i^{\bsa} - \phi_j^{\bsb}),
    \label{eq: lambda defn}
\end{equation}
agree with prescription Eq.~(\ref{eq:NetwCurrsNew}). In this formulation, mutual currents correspond to the symmetry \smash{$\Lambda^{\bsa \bsb}_{ij}=\Lambda^{\bsb\bsa }_{ji}$} and non-mutual current flows arise across inductors for which this symmetry is broken.  As in the historical analogy, passive springs ($k_{\text{p}}$) are encoded by passive inductors which contribute inverse inductances
\begin{equation}
    \Lambda_{ij}^{\bsa, \bsa \pm \mathbf e_i} = k_{\text p} \delta_{ij}.
    \label{eq: lambda passive}
\end{equation}
These contributions, as illustrated by blue arrows in Fig.~\ref{Fig:Circuit}b, are mutual. On the other hand, active springs ($k_{\text{a}}$) contribute terms
\begin{align}    
    \Lambda_{ij}^{\bsa, \bsa \pm \mathbf e_i} &=k_{\text a} \epsilon_{ij}, \label{eq: lambda active 1}\\
    \Lambda_{ij}^{\bsa, \bsa} &= -2 k_{\text a} \epsilon_{ij},
    \label{eq: lambda active 2}
\end{align}
and therefore give rise to non-mutual currents~(green arrows Fig.~\ref{Fig:Circuit}c). Setting the remaining entries of~\smash{$\Lambda^{\bsa\bsb}_{ij}$} to zero, the idealized current-flux relation Eq.~(\ref{eq: lambda defn}) with inverse inductances Eqs.~(\ref{eq: lambda passive})--(\ref{eq: lambda active 2}) recovers the current response~Eq.~(\ref{eq:NetwCurrsNew}).  Crucially, if~\hbox{$k_{\text a}\ne0$} then some of the inductances described by Eqs.~(\ref{eq: lambda active 1}) and~(\ref{eq: lambda active 2}) are negative, which implies -- even without knowledge about the underlying mechanical analogy -- the need for active circuit elements.

The inductance matrix~\smash{$\Lambda^{\bsa\bsb}_{ij}$} found above reveals the need for non-mutual current flows and therefore implies an apparent violation of Kirchhoff's law. In practice, this is realized by a controlled injection of currents from external sources. We implement active circuit elements constructed around operational amplifiers (OAs)~(Fig.~\ref{Fig:Circuit}e), which inject such external currents autonomously according to flux differences between connected nodes. In the limit of ideal circuit elements and a suitable identification of (effective) inductances (see Fig.~\ref{Fig:Circuit}) the resulting circuit exhibits a current-flux response equivalent to Eq.~(\ref{eq:NetwCurrsNew})~[SI]. 

\subsubsection{Linear current-flux response}\label{sec: statics}
In addition to the theoretical properties of ideal OECs, we will demonstrate their robustness under the non-deal effects present in any realistic circuit. To this end, we employ SPICE-based circuit simulations~\cite{Spice,LTspice} throughout the remaining sections, which incorporate circuit components with realistic properties and tolerances to validate the feasibility of OECs as an experimental system.

\begin{figure*}
\includegraphics[width=\textwidth]{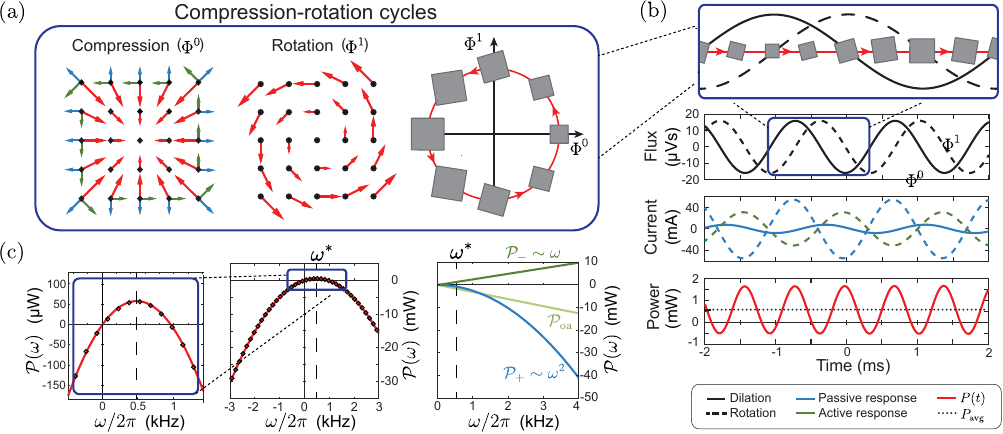}
\caption{Flux cycles and apparent power extraction in simulations of realistic OECs. (a) Static flux patterns (red) analog to dilation ($\Phi^0$) and rotation ($\Phi^1$) give rise to passive (blue) and active (green) currents as predicted by Eq.~(\ref{eq:J Phi coupling}). (b) Cyclical alternation of dilation and rotation profiles [see Eq.~(\ref{eq:phi(t)})] as example of a work-generating cycle. The phase shift between (active) rotational flux ($\Phi^1$) and current (dashed, green), facilitated by non-mutual current flows [see Eq.~(\ref{eq: P avg diss})], gives rise to a non-zero period-averaged power output $P_{\text{avg}}$. (c) Simulation data (black symbols) from a quasi-static frequency sweep (center), showing time-reversal symmetry breaking and a regime of positive power output (left). Fits of time-reversal even ($\mathcal{P}_+$) and time-reversal odd ($\mathcal{P}_-$) components of the measured power show strong agreement (\textless5\% error) with predictions for an ideal circuit [Eq.~(\ref{eq: P omega})] (red line: full fit), including the maximal gain frequency $\omega^*$. Power $\mathcal{P}_{\text{oa}}$ directly measured at OAs reveals the `microscopic" energy source that facilitates the apparent power gain ($\mathcal{P}_{\text{oa}}+\mathcal{P}_-\approx0$).}
\label{fig: work extraction}
\end{figure*}

We first characterize the circuit's current response to different imposed flux profiles, in analogy to performing a mechanical stress-strain measurement. Similar to a strain tensor, we define a flux gradient at site $\bsa$ by 
\begin{equation}
    \Phi_{ij} = \phi_j^{\bsa + \mathbf e_i} - \phi_j^{\bsa}
    \label{eq: gradient phi}
\end{equation}  
and drop site-indices index for brevity. The stress-like counterpart in an OEC is the incident current tensor $\mathcal{I}_{ij}$, representing the current incident at node $(\bsa, j)$ due to its interaction with the neighboring node in the positive $i$-th direction. Using Eqs.~(\ref{eq: lambda defn})--(\ref{eq: lambda active 2}), we~find
\begin{equation}
    \mathcal{I} = \begin{pmatrix}
        k_{\text{p}} \Phi_{xx} & -k_{\text{a}} \Phi_{xx} \\ 
        k_{\text{a}} \Phi_{yy} & k_{\text{p}} \Phi_{yy}
    \end{pmatrix}.
    \label{eq: J components}
\end{equation}
As with force-densities arising from spatial variations of mechanical stress, net currents emerge when~$\mathcal{I}$ varies between lattice sites or at boundaries where $\mathcal{I}\ne0$. Adopting a tensor basis~$\{\tau^\nu\}$ with ~\cite{Scheibner:2020gm,braverman2020topological,fruch23,tan2022}
\begin{equation}
    \begin{aligned}
        \tau^0=\begin{pmatrix}
1 & 0\\
0 & 1
\end{pmatrix},&\hspace{1cm}
\tau^1=\begin{pmatrix}
0 & -1\\
1 & 0
\end{pmatrix},\\
        \tau^2=\begin{pmatrix}
1 & 0\\
0 & -1
\end{pmatrix},&\hspace{1cm}
\tau^3=\begin{pmatrix}
0 & 1\\
1 & 0
\end{pmatrix},
    \end{aligned}
    \label{eq:test}
\end{equation}
we write fluxes and currents, \smash{$\Phi^\nu=\sum_{i,j}\tau^{\nu}_{ij}\Phi_{ij}$} and \smash{$\mathcal{I}^\nu=\sum_{i,j}\tau^{\nu}_{ij}\mathcal{I}_{ij}$}, respectively, in terms of isotropic \hbox{($\nu=0$)}, rotational \hbox{($\nu=1$)} and shear-like \hbox{($\nu=2,3$)} components. Expressing Eq.~(\ref{eq: J components}) in this basis yields
\begin{equation}\label{eq:J Phi coupling}
\begin{pmatrix}
\mathcal{I}^0\\
\mathcal{I}^1
\end{pmatrix}=
\begin{pmatrix}
k_{\text{p}}\\
k_{\text{a}} 
\end{pmatrix}\Phi^0,
\;\;
\begin{pmatrix}
\mathcal{I}^2\\
\mathcal{I}^3    
\end{pmatrix}=
\begin{pmatrix}
k_{\text{p}}\\
-k_{\text{a}}
\end{pmatrix}
\Phi^2,
\end{equation}
which defines the electrical analog of a modulus tensor $C_{\mu \nu}$ by $\mathcal{I}^\mu = \sum_\nu C_{\mu \nu} \Phi^\nu$. An equivalent elastic modulus tensor can be obtained by coarse-graining elastic interactions Eq.~(\ref{eq:NetwForces}) using a conventional dynamical matrix approach~\cite{ashcroft1976solid} (see SI). Antisymmetric contributions to $C_{\mu \nu}$ are proportional to active inductances $k_{\text{a}}$ and signify the loss of reciprocity in the system~\cite{Scheibner:2020gm,fruch23}. Snapshots of the response to flux profiles $\nu=0,1$, obtained from realistic circuit simulations, are shown in Fig.~\ref{fig: work extraction}a (red arrows). As predicted by (\ref{eq:J Phi coupling}), passive currents (blue) have the same symmetry as the imposed dilation-like flux profile ($\Phi^0$), while active currents (green) respond via a rotational pattern ($\mathcal{I}^1$). In turn, a flux profile with rotational character ($\Phi^1$) does not lead to net currents -- an explicit manifestation of non-reciprocity in the circuit~response.

\subsubsection{Apparent power extraction}\label{sec: work extraction}

\begin{figure*}[t]
    \includegraphics[width=\textwidth]{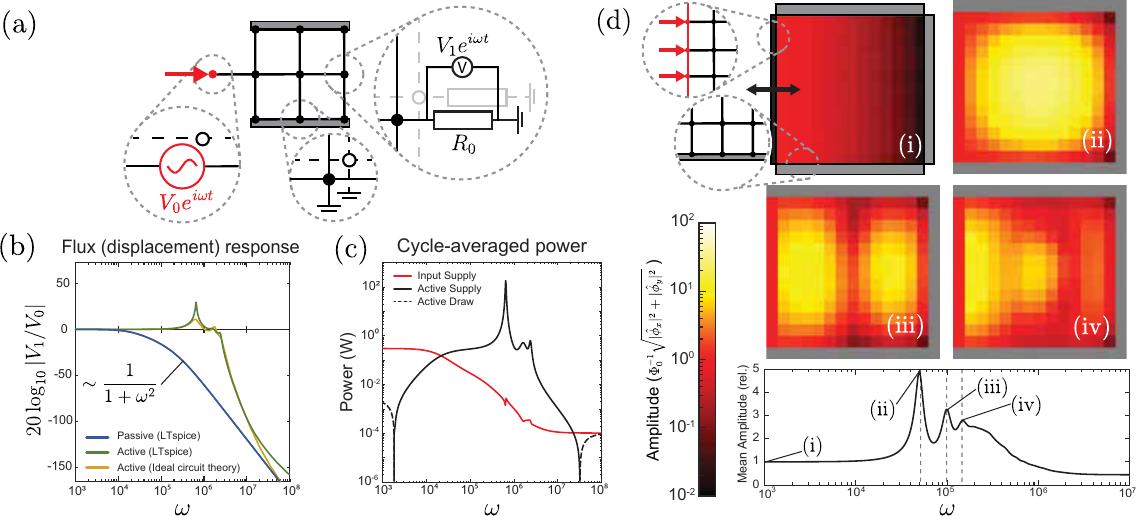}
    \caption{Amplification, resonance and piston-analog excitations in \textit{in-silico} experiments. (a)~Minimal setup to demonstrate amplification and resonance properties of OECs. A 3$\times$3 OEC lattice is grounded at `walls' and receives an AC input via a driving node (red). (b)~Amplification of target node flux ($V_1/i\omega$) shows strong agreement with ideal circuit theory and sustains signal over a frequency range two orders of magnitude larger than a corresponding passive circuit. (c)~Power supply from external driving node (red) is negligible at resonance, where amplification is instead facilitated by active circuit elements (black). (d) Extension of (a) to the electrical analog of an oscillatory piston compression on a 15$\times$15 lattice reveals a sequence of resonances with amplification patterns given by spatial harmonics [SI]. Panels~(ii),~(iii) and~(iv) show the most prominent resonances. Analytically predicted collective resonance frequencies (bottom panel, dashed lines, [SI]) show very good agreement with realistic circuit simulations (solid line).}
   \label{fig: amplification and resonance combined}
\end{figure*}

A hallmark property of non-reciprocal systems is that work, and consequently power, can be extracted by applying periodic driving~\cite{Scheibner:2020gm,Sousl2021,fruch23}. This can of course only be an apparent effect that requires internal energy sources. These sources are typically implicit in the formulation of theoretical models~\cite{Scheibner:2020gm,Sousl2021,Lin2024} or emerge from complex processes, for example, in living systems~\cite{tan2022,Shankar2024}. In contrast, OECs supply ``microscopic" energy via a well-controlled mechanism in which OAs facilitate non-mutual current flows. To exploit the tractability of this mechanism in cyclic excitations of an OEC, we connect each OEC node to ground via a resistive element of conductance $\sigma$, which corresponds to adding the electrical analog of a substrate friction. In this case, the circuit response Eq.~(\ref{eq: lambda defn}) generalizes  to the damped dynamics,
\begin{equation}
    I_i^{\bsa} = -\sum_{j,\bsb}\tilde\Lambda_{ij}^{\bsa \bsb} \phi_j^{\bsb} - \sigma V_i^{\bsa},
    \label{eq: currents w damping}
\end{equation}
where, for brevity, we introduce the modified inverse inductance matrix,
\begin{equation} \label{eq: lambda tilde}
\tilde\Lambda_{ij}^{\bsa \bsb} = \Lambda_{ij}^{\bsa \bsb} - \left( \textstyle{\sum_{k,\boldsymbol{\gamma}}} \Lambda_{ik}^{\bsa\boldsymbol{\gamma}} \right) \delta_{ij} \delta^{\bsa\bsb}.
\end{equation}
The net instantaneous power output of the OEC is then given by
\begin{equation}
    P(t) = - V_i^{\bsa} \tilde\Lambda_{ij}^{\bsa \bsb} \phi_j^{\bsb}- \sigma V_i^{\bsa} V_i^{\bsa},
    \label{eq: P sum}
\end{equation}
where repeated indices now indicate summation. The time averaged power output per lattice site under a cyclic excitation of duration $T$ therefore has contributions from inductive and resistive circuit elements given by
\begin{align}
    \mathcal P_{\text{ind}}&= \frac{1}{TN^2}\oint dt \left[ \frac{1}{2}V_i^{\bsa} (\Lambda_{ji}^{\bsb \bsa}-\Lambda_{ij}^{\bsa \bsb})\phi_j^{\bsb} \right]\label{eq:P avg el}\\
    \mathcal P_{\text{res}} &= \frac{1}{TN^2}\oint dt \left[  - \sigma V_i^{\bsa} V_i^{\bsa} \right].    
    \label{eq: P avg diss}
\end{align}
and the total power output per lattice site is~\hbox{$\mathcal P=\mathcal P_{\text{ind}}+\mathcal P_{\text{res}}$}. Equation~(\ref{eq:P avg el}) shows that non-mutual current flows, described by the antisymmetric part of~\smash{$\Lambda_{ij}^{\bsa \bsb}$}, may in principle lead to a power $\mathcal P_{\text{ind}}>0$ that could be extracted from an OEC. 

In analogy to an odd elastic solid~\cite{Scheibner:2020gm}, one can demonstrate this apparent power extraction by applying time-periodic flux gradients of the form
\begin{equation}
    \label{eq:phi(t)}
    \Phi_0^{-1} \Phi_{ij}(t) = \cos(\omega t) \tau_{ij}^\nu + \sin(\omega t) \tau_{ij}^\mu,
\end{equation}
where  $\Phi_0$ is a fixed amplitude. We consider the electrical encodings of a rotation-compression cycle ($\nu=1,\mu=0$)~(Fig.~\ref{fig: work extraction}a,b) and a shear-shear cycle ($\nu=2,\mu=3$). It then follows from Eqs.~(\ref{eq:P avg el}), (\ref{eq: P avg diss}) that the total power output $\mathcal P(\omega)$ is given by~[SI]
\begin{equation}\label{eq: P omega}
\mathcal{P}(\omega)= \Phi_0^2\left[(1-N^{-1})k_{\text{a}}\omega - c_{N}\sigma\omega^2\right].
\end{equation}
As expected for friction-analog contributions in a finite system, the positive constant \hbox{$c_{N}\ge(N^2-1)/6$} depends on the location of the grounded reference site $(V_i^{\bsa}=0)$ at which flux patterns (\ref{eq:phi(t)}) are centered. A finite amount of extractable work remains even in the adiabatic limit $\omega\rightarrow0$ [SI], such that OECs are a realization of optimal active engines~\cite{Sousl2021,etien2021b}. 

To validate Eq.~(\ref{eq: P omega}) in realistic circuit simulations, we perform quasi-static frequency sweeps on a \hbox{$5\times5$} lattice to obtain data for $\mathcal P(\omega)$ (Fig.~\ref{fig: work extraction}c). As predicted by Eq.~(\ref{eq: P omega}), we observe a finite frequency band in which power is extracted from the OEC ($\mathcal{P}>0$) and quantitatively recover the predicted frequency $\omega^*=3k_{\text{a}}/[\sigma N(N+1)]$ of maximal gain (dashed line) within a 5\,\% error. Decomposing the extracted power into even and odd parts, $\mathcal{P}_+=[\mathcal{P}(\omega)+\mathcal{P}(-\omega)]/2$ and $\mathcal{P}_-=[\mathcal{P}(\omega)-\mathcal{P}(-\omega)]/2$, respectively, isolates inductive (odd) and resistive (even) contributions, according to Eq.~(\ref{eq: P omega}). From this, we infer the ``microscopic" circuit parameters $k_{\text{a}}$ and $\sigma$ directly from power extraction measurements of realistic circuit simulations, again within an error of less than~5\,\%. Finally, we exploit our access to the hidden internal degrees of freedom in OECs and determine the time-odd part of the power needed to operate OAs during cycles ($\mathcal{P}_{\text{oa}}$ in Fig.~2c). This quantitatively illustrates how OAs serve as `microscopic" energy sources that facilitate the apparent power gain ($\mathcal{P}_{\text{oa}}+\mathcal{P}_-\approx0$) up to a loss of approximately~$20\%$.

\subsubsection{Amplification and resonance in OECs}

Finally, we exploit the scalability of OECs to investigate their transmission properties when treated as a lumped element. A minimal example of such a transmission test is shown in Fig.~\ref{fig: amplification and resonance combined}a for a 3$\times$3 lattice, where an AC voltage source (magnitude $|V_0|$) excites a single $x$-node at the boundary of the circuit. We find in both the ideal circuit theory and realistic circuit simulations that the received signal $\sim|V_1|$ is sustained over a frequency band that is wider by more than two orders of magnitude than for an otherwise equivalent passive ($k_{\text{a}}=0$) circuit (Fig.~\ref{fig: amplification and resonance combined}b), achieving amplifications of almost 30dB in a narrow band. An external and internal power balance similar to Fig.~\ref{fig: work extraction}c shows that the external driving node makes negligible power contributions at large gain frequencies, where energy is instead supplied by internal active circuit elements (Fig.~\ref{fig: amplification and resonance combined}c).

We rationalize this amplification as a resonance between the periodic boundary driving and active circuit elements that counteract damping. To this end, we utilize the scalability of OECs and demonstrate a sequence of resonances under piston-analog compression (Fig.~\ref{fig: amplification and resonance combined}d). Specifically, we consider a 15$\times 15$ OEC lattice and an external AC signal supply along one edge, while all other boundary nodes remain grounded -- the electrical analog of a clamped odd elastic material under unidirectional oscillatory compression. Despite the grounding of most boundary nodes, this system still exhibits a broad frequency band in which the oscillatory input amplitude is amplified. Using ideal circuit theory, emerging amplification patterns at different resonances (Fig.~\ref{fig: amplification and resonance combined}d, top) can be understood in terms of spatial harmonics~[SI]. Corresponding analytic predictions for collective resonance frequencies (black dashed lines) show strong agreement with realistic circuit simulations~[SI]. Identifying this amplification, even in the presence of strong damping, suggests a thus far unexplored functionality of odd elastic materials as robust mechanical amplifiers. 

\section{Conclusion}

We have introduced odd electrical circuits (OECs), a novel class of electrical circuits designed to emulate odd elastic materials. By integrating active circuit elements~\cite{kotwal21}, OECs attain non-reciprocal properties and exhibit a range of hallmark "odd" features. Through SPICE-based simulations with realistic circuit elements, we demonstrate the robustness of these properties, establishing OECs as a versatile model system for studying the associated class of non-equilibrium behaviors. Leveraging computational tools for AC circuit analysis, we have identified boundary forcing regimes that enable tunable active resonances, potentially providing an efficient mechanism for work extraction. Moreover, the nature of the observed resonances is to amplify global chiral oscillations, qualitatively resembling those recently observed in experiments involving living chiral crystals~\cite{tan2022,chao2024}. We anticipate that incorporating nonlinear components will also allow the realization of limit cycle behaviors similar to those observed in these experiments. 

Building on previous research in electronically mediated activity in mechanical systems~\cite{Chen2021,nash2015topological,wang2021emergent}, OECs offer a fully electrical, scalable experimental platform for studying active metamaterials. Additional avenues for future exploration include manipulating spatial structures to create topologically non-trivial circuits and investigating the impact of chiral symmetry breaking on OECs as receivers.

\begin{acknowledgments}
We thank Anton Souslov for helpful discussions. This work was supported by a MathWorks Science Fellowship (H.W.), the MathWorks Professorship Fund (J.D.), Alfred P. Sloan Foundation Grant G-2021-16758 (J.D.), and through Schmidt Sciences LLC (Polymath award to J.D.).
\end{acknowledgments}


\begin{thebibliography}{34}%
\makeatletter
\providecommand \@ifxundefined [1]{%
 \@ifx{#1\undefined}
}%
\providecommand \@ifnum [1]{%
 \ifnum #1\expandafter \@firstoftwo
 \else \expandafter \@secondoftwo
 \fi
}%
\providecommand \@ifx [1]{%
 \ifx #1\expandafter \@firstoftwo
 \else \expandafter \@secondoftwo
 \fi
}%
\providecommand \natexlab [1]{#1}%
\providecommand \enquote  [1]{``#1''}%
\providecommand \bibnamefont  [1]{#1}%
\providecommand \bibfnamefont [1]{#1}%
\providecommand \citenamefont [1]{#1}%
\providecommand \href@noop [0]{\@secondoftwo}%
\providecommand \href [0]{\begingroup \@sanitize@url \@href}%
\providecommand \@href[1]{\@@startlink{#1}\@@href}%
\providecommand \@@href[1]{\endgroup#1\@@endlink}%
\providecommand \@sanitize@url [0]{\catcode `\\12\catcode `\$12\catcode
  `\&12\catcode `\#12\catcode `\^12\catcode `\_12\catcode `\%12\relax}%
\providecommand \@@startlink[1]{}%
\providecommand \@@endlink[0]{}%
\providecommand \url  [0]{\begingroup\@sanitize@url \@url }%
\providecommand \@url [1]{\endgroup\@href {#1}{\urlprefix }}%
\providecommand \urlprefix  [0]{URL }%
\providecommand \Eprint [0]{\href }%
\providecommand \doibase [0]{https://doi.org/}%
\providecommand \selectlanguage [0]{\@gobble}%
\providecommand \bibinfo  [0]{\@secondoftwo}%
\providecommand \bibfield  [0]{\@secondoftwo}%
\providecommand \translation [1]{[#1]}%
\providecommand \BibitemOpen [0]{}%
\providecommand \bibitemStop [0]{}%
\providecommand \bibitemNoStop [0]{.\EOS\space}%
\providecommand \EOS [0]{\spacefactor3000\relax}%
\providecommand \BibitemShut  [1]{\csname bibitem#1\endcsname}%
\let\auto@bib@innerbib\@empty
\bibitem [{\citenamefont {Scheibner}\ \emph
  {et~al.}(2020{\natexlab{a}})\citenamefont {Scheibner}, \citenamefont
  {Souslov}, \citenamefont {Banerjee}, \citenamefont {Sur{\'o}wka},
  \citenamefont {Irvine},\ and\ \citenamefont {Vitelli}}]{Scheibner:2020gm}%
  \BibitemOpen
  \bibfield  {author} {\bibinfo {author} {\bibfnamefont {C.}~\bibnamefont
  {Scheibner}}, \bibinfo {author} {\bibfnamefont {A.}~\bibnamefont {Souslov}},
  \bibinfo {author} {\bibfnamefont {D.}~\bibnamefont {Banerjee}}, \bibinfo
  {author} {\bibfnamefont {P.}~\bibnamefont {Sur{\'o}wka}}, \bibinfo {author}
  {\bibfnamefont {W.~T.~M.}\ \bibnamefont {Irvine}},\ and\ \bibinfo {author}
  {\bibfnamefont {V.}~\bibnamefont {Vitelli}},\ }\href@noop {} {\bibfield
  {journal} {\bibinfo  {journal} {Nat. Phys.}\ }\textbf {\bibinfo {volume}
  {16}},\ \bibinfo {pages} {475} (\bibinfo {year}
  {2020}{\natexlab{a}})}\BibitemShut {NoStop}%
\bibitem [{\citenamefont {Fruchart}\ \emph
  {et~al.}(2023{\natexlab{a}})\citenamefont {Fruchart}, \citenamefont
  {Scheibner},\ and\ \citenamefont {Vitelli}}]{fruch23rev}%
  \BibitemOpen
  \bibfield  {author} {\bibinfo {author} {\bibfnamefont {M.}~\bibnamefont
  {Fruchart}}, \bibinfo {author} {\bibfnamefont {C.}~\bibnamefont
  {Scheibner}},\ and\ \bibinfo {author} {\bibfnamefont {V.}~\bibnamefont
  {Vitelli}},\ }\href@noop {} {\bibfield  {journal} {\bibinfo  {journal} {Annu.
  Rev. Condens. Matter Phys.}\ }\textbf {\bibinfo {volume} {14}},\ \bibinfo
  {pages} {471} (\bibinfo {year} {2023}{\natexlab{a}})}\BibitemShut {NoStop}%
\bibitem [{\citenamefont {Coulais}\ \emph {et~al.}(2017)\citenamefont
  {Coulais}, \citenamefont {Sounas},\ and\ \citenamefont {Al{\`u}}}]{coul17}%
  \BibitemOpen
  \bibfield  {author} {\bibinfo {author} {\bibfnamefont {C.}~\bibnamefont
  {Coulais}}, \bibinfo {author} {\bibfnamefont {D.}~\bibnamefont {Sounas}},\
  and\ \bibinfo {author} {\bibfnamefont {A.}~\bibnamefont {Al{\`u}}},\
  }\href@noop {} {\bibfield  {journal} {\bibinfo  {journal} {Nature}\ }\textbf
  {\bibinfo {volume} {542}},\ \bibinfo {pages} {461} (\bibinfo {year}
  {2017})}\BibitemShut {NoStop}%
\bibitem [{\citenamefont {Scheibner}\ \emph
  {et~al.}(2020{\natexlab{b}})\citenamefont {Scheibner}, \citenamefont
  {Irvine},\ and\ \citenamefont {Vitelli}}]{scheib20}%
  \BibitemOpen
  \bibfield  {author} {\bibinfo {author} {\bibfnamefont {C.}~\bibnamefont
  {Scheibner}}, \bibinfo {author} {\bibfnamefont {W.~T.~M.}\ \bibnamefont
  {Irvine}},\ and\ \bibinfo {author} {\bibfnamefont {V.}~\bibnamefont
  {Vitelli}},\ }\href@noop {} {\bibfield  {journal} {\bibinfo  {journal} {Phys.
  Rev. Lett.}\ }\textbf {\bibinfo {volume} {125}},\ \bibinfo {pages} {118001}
  (\bibinfo {year} {2020}{\natexlab{b}})}\BibitemShut {NoStop}%
\bibitem [{\citenamefont {Fossati}\ \emph {et~al.}(2024)\citenamefont
  {Fossati}, \citenamefont {Scheibner}, \citenamefont {Fruchart},\ and\
  \citenamefont {Vitelli}}]{Fossati2024}%
  \BibitemOpen
  \bibfield  {author} {\bibinfo {author} {\bibfnamefont {M.}~\bibnamefont
  {Fossati}}, \bibinfo {author} {\bibfnamefont {C.}~\bibnamefont {Scheibner}},
  \bibinfo {author} {\bibfnamefont {M.}~\bibnamefont {Fruchart}},\ and\
  \bibinfo {author} {\bibfnamefont {V.}~\bibnamefont {Vitelli}},\ }\href
  {https://doi.org/10.1103/PhysRevE.109.024608} {\bibfield  {journal} {\bibinfo
   {journal} {Phys. Rev. E}\ }\textbf {\bibinfo {volume} {109}},\ \bibinfo
  {pages} {024608} (\bibinfo {year} {2024})}\BibitemShut {NoStop}%
\bibitem [{\citenamefont {Braverman}\ \emph {et~al.}(2021)\citenamefont
  {Braverman}, \citenamefont {Scheibner}, \citenamefont {VanSaders},\ and\
  \citenamefont {Vitelli}}]{braverman2020topological}%
  \BibitemOpen
  \bibfield  {author} {\bibinfo {author} {\bibfnamefont {L.}~\bibnamefont
  {Braverman}}, \bibinfo {author} {\bibfnamefont {C.}~\bibnamefont
  {Scheibner}}, \bibinfo {author} {\bibfnamefont {B.}~\bibnamefont
  {VanSaders}},\ and\ \bibinfo {author} {\bibfnamefont {V.}~\bibnamefont
  {Vitelli}},\ }\href@noop {} {\bibfield  {journal} {\bibinfo  {journal} {Phys.
  Rev. Lett.}\ }\textbf {\bibinfo {volume} {127}},\ \bibinfo {pages} {268001}
  (\bibinfo {year} {2021})}\BibitemShut {NoStop}%
\bibitem [{\citenamefont {Poncet}\ and\ \citenamefont
  {Bartolo}(2022)}]{ponc22}%
  \BibitemOpen
  \bibfield  {author} {\bibinfo {author} {\bibfnamefont {A.}~\bibnamefont
  {Poncet}}\ and\ \bibinfo {author} {\bibfnamefont {D.}~\bibnamefont
  {Bartolo}},\ }\href@noop {} {\bibfield  {journal} {\bibinfo  {journal} {Phys.
  Rev. Lett.}\ }\textbf {\bibinfo {volume} {128}},\ \bibinfo {pages} {048002}
  (\bibinfo {year} {2022})}\BibitemShut {NoStop}%
\bibitem [{\citenamefont {Truesdell}(1963)}]{true63}%
  \BibitemOpen
  \bibfield  {author} {\bibinfo {author} {\bibfnamefont {C.~A.}\ \bibnamefont
  {Truesdell}},\ }\href@noop {} {\bibfield  {journal} {\bibinfo  {journal} {J.
  Res. Natl. Bur. Stand. B}\ }\textbf {\bibinfo {volume} {67B}},\ \bibinfo
  {pages} {85} (\bibinfo {year} {1963})}\BibitemShut {NoStop}%
\bibitem [{\citenamefont {Fodor}\ and\ \citenamefont
  {Cates}(2021)}]{etien2021b}%
  \BibitemOpen
  \bibfield  {author} {\bibinfo {author} {\bibfnamefont {{\'{E}}.}~\bibnamefont
  {Fodor}}\ and\ \bibinfo {author} {\bibfnamefont {M.~E.}\ \bibnamefont
  {Cates}},\ }\href@noop {} {\bibfield  {journal} {\bibinfo  {journal} {EPL}\
  }\textbf {\bibinfo {volume} {134}},\ \bibinfo {pages} {10003} (\bibinfo
  {year} {2021})}\BibitemShut {NoStop}%
\bibitem [{\citenamefont {Fodor}\ and\ \citenamefont
  {Souslov}(2021)}]{Sousl2021}%
  \BibitemOpen
  \bibfield  {author} {\bibinfo {author} {\bibfnamefont {{\'{E}}.}~\bibnamefont
  {Fodor}}\ and\ \bibinfo {author} {\bibfnamefont {A.}~\bibnamefont
  {Souslov}},\ }\href@noop {} {\bibfield  {journal} {\bibinfo  {journal} {Phys.
  Rev. E}\ }\textbf {\bibinfo {volume} {104}},\ \bibinfo {pages} {L062602}
  (\bibinfo {year} {2021})}\BibitemShut {NoStop}%
\bibitem [{\citenamefont {Kobayashi}\ \emph {et~al.}(2023)\citenamefont
  {Kobayashi}, \citenamefont {Yasuda}, \citenamefont {Ishimoto}, \citenamefont
  {Lin}, \citenamefont {Sou}, \citenamefont {Hosaka},\ and\ \citenamefont
  {Komura}}]{koba23}%
  \BibitemOpen
  \bibfield  {author} {\bibinfo {author} {\bibfnamefont {A.}~\bibnamefont
  {Kobayashi}}, \bibinfo {author} {\bibfnamefont {K.}~\bibnamefont {Yasuda}},
  \bibinfo {author} {\bibfnamefont {K.}~\bibnamefont {Ishimoto}}, \bibinfo
  {author} {\bibfnamefont {L.-S.}\ \bibnamefont {Lin}}, \bibinfo {author}
  {\bibfnamefont {I.}~\bibnamefont {Sou}}, \bibinfo {author} {\bibfnamefont
  {Y.}~\bibnamefont {Hosaka}},\ and\ \bibinfo {author} {\bibfnamefont
  {S.}~\bibnamefont {Komura}},\ }\href@noop {} {\bibfield  {journal} {\bibinfo
  {journal} {J. Phys. Soc. Jpn.}\ }\textbf {\bibinfo {volume} {92}},\ \bibinfo
  {pages} {074801} (\bibinfo {year} {2023})}\BibitemShut {NoStop}%
\bibitem [{\citenamefont {Cheng}\ and\ \citenamefont {Hu}(2021)}]{cheng21}%
  \BibitemOpen
  \bibfield  {author} {\bibinfo {author} {\bibfnamefont {W.}~\bibnamefont
  {Cheng}}\ and\ \bibinfo {author} {\bibfnamefont {G.}~\bibnamefont {Hu}},\
  }\href@noop {} {\bibfield  {journal} {\bibinfo  {journal} {Science China
  Physics, Mechanics \& Astronomy}\ }\textbf {\bibinfo {volume} {64}} (\bibinfo
  {year} {2021})}\BibitemShut {NoStop}%
\bibitem [{\citenamefont {Brandenbourger}\ \emph {et~al.}(2019)\citenamefont
  {Brandenbourger}, \citenamefont {Locsin}, \citenamefont {Lerner},\ and\
  \citenamefont {Coulais}}]{Brandenbourger2019}%
  \BibitemOpen
  \bibfield  {author} {\bibinfo {author} {\bibfnamefont {M.}~\bibnamefont
  {Brandenbourger}}, \bibinfo {author} {\bibfnamefont {X.}~\bibnamefont
  {Locsin}}, \bibinfo {author} {\bibfnamefont {E.}~\bibnamefont {Lerner}},\
  and\ \bibinfo {author} {\bibfnamefont {C.}~\bibnamefont {Coulais}},\
  }\href@noop {} {\bibfield  {journal} {\bibinfo  {journal} {Nat. Commun.}\
  }\textbf {\bibinfo {volume} {10}},\ \bibinfo {pages} {4608} (\bibinfo {year}
  {2019})}\BibitemShut {NoStop}%
\bibitem [{\citenamefont {Chen}\ \emph {et~al.}(2021)\citenamefont {Chen},
  \citenamefont {Li}, \citenamefont {Scheibner}, \citenamefont {Vitelli},\ and\
  \citenamefont {Huang}}]{Chen2021}%
  \BibitemOpen
  \bibfield  {author} {\bibinfo {author} {\bibfnamefont {Y.}~\bibnamefont
  {Chen}}, \bibinfo {author} {\bibfnamefont {X.}~\bibnamefont {Li}}, \bibinfo
  {author} {\bibfnamefont {C.}~\bibnamefont {Scheibner}}, \bibinfo {author}
  {\bibfnamefont {V.}~\bibnamefont {Vitelli}},\ and\ \bibinfo {author}
  {\bibfnamefont {G.}~\bibnamefont {Huang}},\ }\href
  {https://doi.org/10.1038/s41467-021-26034-z} {\bibfield  {journal} {\bibinfo
  {journal} {Nat. Commun.}\ }\textbf {\bibinfo {volume} {12}},\ \bibinfo
  {pages} {5935} (\bibinfo {year} {2021})}\BibitemShut {NoStop}%
\bibitem [{\citenamefont {Brandenbourger}\ \emph {et~al.}(2022)\citenamefont
  {Brandenbourger}, \citenamefont {Scheibner}, \citenamefont {Veenstra},
  \citenamefont {Vitelli},\ and\ \citenamefont {Coulais}}]{brandenbourger2022}%
  \BibitemOpen
  \bibfield  {author} {\bibinfo {author} {\bibfnamefont {M.}~\bibnamefont
  {Brandenbourger}}, \bibinfo {author} {\bibfnamefont {C.}~\bibnamefont
  {Scheibner}}, \bibinfo {author} {\bibfnamefont {J.}~\bibnamefont {Veenstra}},
  \bibinfo {author} {\bibfnamefont {V.}~\bibnamefont {Vitelli}},\ and\ \bibinfo
  {author} {\bibfnamefont {C.}~\bibnamefont {Coulais}},\ }\href
  {https://arxiv.org/abs/2108.08837} {\bibinfo {title} {Limit cycles turn
  active matter into robots}} (\bibinfo {year} {2022}),\ \Eprint
  {https://arxiv.org/abs/2108.08837} {arXiv:2108.08837 [cond-mat.soft]}
  \BibitemShut {NoStop}%
\bibitem [{\citenamefont {Bililign}\ \emph {et~al.}(2022)\citenamefont
  {Bililign}, \citenamefont {Balboa~Usabiaga}, \citenamefont {Ganan},
  \citenamefont {Poncet}, \citenamefont {Soni}, \citenamefont {Magkiriadou},
  \citenamefont {Shelley}, \citenamefont {Bartolo},\ and\ \citenamefont
  {Irvine}}]{bili21}%
  \BibitemOpen
  \bibfield  {author} {\bibinfo {author} {\bibfnamefont {E.~S.}\ \bibnamefont
  {Bililign}}, \bibinfo {author} {\bibfnamefont {F.}~\bibnamefont
  {Balboa~Usabiaga}}, \bibinfo {author} {\bibfnamefont {Y.~A.}\ \bibnamefont
  {Ganan}}, \bibinfo {author} {\bibfnamefont {A.}~\bibnamefont {Poncet}},
  \bibinfo {author} {\bibfnamefont {V.}~\bibnamefont {Soni}}, \bibinfo {author}
  {\bibfnamefont {S.}~\bibnamefont {Magkiriadou}}, \bibinfo {author}
  {\bibfnamefont {M.~J.}\ \bibnamefont {Shelley}}, \bibinfo {author}
  {\bibfnamefont {D.}~\bibnamefont {Bartolo}},\ and\ \bibinfo {author}
  {\bibfnamefont {W.~T.~M.}\ \bibnamefont {Irvine}},\ }\href@noop {} {\bibfield
   {journal} {\bibinfo  {journal} {Nat. Phys.}\ }\textbf {\bibinfo {volume}
  {18}},\ \bibinfo {pages} {212} (\bibinfo {year} {2022})}\BibitemShut
  {NoStop}%
\bibitem [{\citenamefont {Tan}\ \emph {et~al.}(2022)\citenamefont {Tan},
  \citenamefont {Mietke}, \citenamefont {Li}, \citenamefont {Chen},
  \citenamefont {Higinbotham}, \citenamefont {Foster}, \citenamefont {Gokhale},
  \citenamefont {Dunkel},\ and\ \citenamefont {Fakhri}}]{tan2022}%
  \BibitemOpen
  \bibfield  {author} {\bibinfo {author} {\bibfnamefont {T.~H.}\ \bibnamefont
  {Tan}}, \bibinfo {author} {\bibfnamefont {A.}~\bibnamefont {Mietke}},
  \bibinfo {author} {\bibfnamefont {J.}~\bibnamefont {Li}}, \bibinfo {author}
  {\bibfnamefont {Y.}~\bibnamefont {Chen}}, \bibinfo {author} {\bibfnamefont
  {H.}~\bibnamefont {Higinbotham}}, \bibinfo {author} {\bibfnamefont {P.~J.}\
  \bibnamefont {Foster}}, \bibinfo {author} {\bibfnamefont {S.}~\bibnamefont
  {Gokhale}}, \bibinfo {author} {\bibfnamefont {J.}~\bibnamefont {Dunkel}},\
  and\ \bibinfo {author} {\bibfnamefont {N.}~\bibnamefont {Fakhri}},\ }\href
  {https://doi.org/10.1038/s41586-022-04889-6} {\bibfield  {journal} {\bibinfo
  {journal} {Nature}\ }\textbf {\bibinfo {volume} {607}},\ \bibinfo {pages}
  {287} (\bibinfo {year} {2022})}\BibitemShut {NoStop}%
\bibitem [{\citenamefont {Ishimoto}\ \emph {et~al.}(2023)\citenamefont
  {Ishimoto}, \citenamefont {Moreau},\ and\ \citenamefont
  {Yasuda}}]{Ishimoto2023}%
  \BibitemOpen
  \bibfield  {author} {\bibinfo {author} {\bibfnamefont {K.}~\bibnamefont
  {Ishimoto}}, \bibinfo {author} {\bibfnamefont {C.}~\bibnamefont {Moreau}},\
  and\ \bibinfo {author} {\bibfnamefont {K.}~\bibnamefont {Yasuda}},\ }\href
  {https://doi.org/10.1103/PRXLife.1.023002} {\bibfield  {journal} {\bibinfo
  {journal} {PRX Life}\ }\textbf {\bibinfo {volume} {1}},\ \bibinfo {pages}
  {023002} (\bibinfo {year} {2023})}\BibitemShut {NoStop}%
\bibitem [{\citenamefont {Shankar}\ and\ \citenamefont
  {Mahadevan}(2024)}]{Shankar2024}%
  \BibitemOpen
  \bibfield  {author} {\bibinfo {author} {\bibfnamefont {S.}~\bibnamefont
  {Shankar}}\ and\ \bibinfo {author} {\bibfnamefont {L.}~\bibnamefont
  {Mahadevan}},\ }\href {https://doi.org/10.1038/s41567-024-02540-x} {\bibfield
   {journal} {\bibinfo  {journal} {Nat. Phys.}\ }\textbf {\bibinfo {volume}
  {20}},\ \bibinfo {pages} {1501} (\bibinfo {year} {2024})}\BibitemShut
  {NoStop}%
\bibitem [{\citenamefont {Sur\'owka}\ \emph {et~al.}(2023)\citenamefont
  {Sur\'owka}, \citenamefont {Souslov}, \citenamefont {J\"ulicher},\ and\
  \citenamefont {Banerjee}}]{Surowka2023}%
  \BibitemOpen
  \bibfield  {author} {\bibinfo {author} {\bibfnamefont {P.}~\bibnamefont
  {Sur\'owka}}, \bibinfo {author} {\bibfnamefont {A.}~\bibnamefont {Souslov}},
  \bibinfo {author} {\bibfnamefont {F.}~\bibnamefont {J\"ulicher}},\ and\
  \bibinfo {author} {\bibfnamefont {D.}~\bibnamefont {Banerjee}},\ }\href
  {https://doi.org/10.1103/PhysRevE.108.064609} {\bibfield  {journal} {\bibinfo
   {journal} {Phys. Rev. E}\ }\textbf {\bibinfo {volume} {108}},\ \bibinfo
  {pages} {064609} (\bibinfo {year} {2023})}\BibitemShut {NoStop}%
\bibitem [{\citenamefont {Olson}(1944)}]{olson44}%
  \BibitemOpen
  \bibfield  {author} {\bibinfo {author} {\bibfnamefont {H.}~\bibnamefont
  {Olson}},\ }\href@noop {} {\emph {\bibinfo {title} {Dynamical Analogies}}}\
  (\bibinfo  {publisher} {D. Van Nostrand},\ \bibinfo {year}
  {1944})\BibitemShut {NoStop}%
\bibitem [{\citenamefont {Firestone}(1933)}]{fire33}%
  \BibitemOpen
  \bibfield  {author} {\bibinfo {author} {\bibfnamefont {F.~A.}\ \bibnamefont
  {Firestone}},\ }\href@noop {} {\bibfield  {journal} {\bibinfo  {journal} {J.
  Acoust. Soc. Am.}\ }\textbf {\bibinfo {volume} {4}},\ \bibinfo {pages} {249}
  (\bibinfo {year} {1933})}\BibitemShut {NoStop}%
\bibitem [{\citenamefont {Kotwal}\ \emph {et~al.}(2021)\citenamefont {Kotwal},
  \citenamefont {Moseley}, \citenamefont {Stegmaier}, \citenamefont {Imhof},
  \citenamefont {Brand}, \citenamefont {Kießling}, \citenamefont {Thomale},
  \citenamefont {Ronellenfitsch},\ and\ \citenamefont {Dunkel}}]{kotwal21}%
  \BibitemOpen
  \bibfield  {author} {\bibinfo {author} {\bibfnamefont {T.}~\bibnamefont
  {Kotwal}}, \bibinfo {author} {\bibfnamefont {F.}~\bibnamefont {Moseley}},
  \bibinfo {author} {\bibfnamefont {A.}~\bibnamefont {Stegmaier}}, \bibinfo
  {author} {\bibfnamefont {S.}~\bibnamefont {Imhof}}, \bibinfo {author}
  {\bibfnamefont {H.}~\bibnamefont {Brand}}, \bibinfo {author} {\bibfnamefont
  {T.}~\bibnamefont {Kießling}}, \bibinfo {author} {\bibfnamefont
  {R.}~\bibnamefont {Thomale}}, \bibinfo {author} {\bibfnamefont
  {H.}~\bibnamefont {Ronellenfitsch}},\ and\ \bibinfo {author} {\bibfnamefont
  {J.}~\bibnamefont {Dunkel}},\ }\href@noop {} {\bibfield  {journal} {\bibinfo
  {journal} {Proc. Natl. Acad. Sci. U.S.A.}\ }\textbf {\bibinfo {volume}
  {118}},\ \bibinfo {pages} {e2106411118} (\bibinfo {year} {2021})}\BibitemShut
  {NoStop}%
\bibitem [{\citenamefont {Stegmaier}\ \emph {et~al.}(2021)\citenamefont
  {Stegmaier}, \citenamefont {Imhof}, \citenamefont {Helbig}, \citenamefont
  {Hofmann}, \citenamefont {Lee}, \citenamefont {Kremer}, \citenamefont
  {Fritzsche}, \citenamefont {Feichtner}, \citenamefont {Klembt}, \citenamefont
  {H\"ofling}, \citenamefont {Boettcher}, \citenamefont {Fulga}, \citenamefont
  {Ma}, \citenamefont {Schmidt}, \citenamefont {Greiter}, \citenamefont
  {Kiessling}, \citenamefont {Szameit},\ and\ \citenamefont
  {Thomale}}]{steg21}%
  \BibitemOpen
  \bibfield  {author} {\bibinfo {author} {\bibfnamefont {A.}~\bibnamefont
  {Stegmaier}}, \bibinfo {author} {\bibfnamefont {S.}~\bibnamefont {Imhof}},
  \bibinfo {author} {\bibfnamefont {T.}~\bibnamefont {Helbig}}, \bibinfo
  {author} {\bibfnamefont {T.}~\bibnamefont {Hofmann}}, \bibinfo {author}
  {\bibfnamefont {C.~H.}\ \bibnamefont {Lee}}, \bibinfo {author} {\bibfnamefont
  {M.}~\bibnamefont {Kremer}}, \bibinfo {author} {\bibfnamefont
  {A.}~\bibnamefont {Fritzsche}}, \bibinfo {author} {\bibfnamefont
  {T.}~\bibnamefont {Feichtner}}, \bibinfo {author} {\bibfnamefont
  {S.}~\bibnamefont {Klembt}}, \bibinfo {author} {\bibfnamefont
  {S.}~\bibnamefont {H\"ofling}}, \bibinfo {author} {\bibfnamefont
  {I.}~\bibnamefont {Boettcher}}, \bibinfo {author} {\bibfnamefont {I.~C.}\
  \bibnamefont {Fulga}}, \bibinfo {author} {\bibfnamefont {L.}~\bibnamefont
  {Ma}}, \bibinfo {author} {\bibfnamefont {O.~G.}\ \bibnamefont {Schmidt}},
  \bibinfo {author} {\bibfnamefont {M.}~\bibnamefont {Greiter}}, \bibinfo
  {author} {\bibfnamefont {T.}~\bibnamefont {Kiessling}}, \bibinfo {author}
  {\bibfnamefont {A.}~\bibnamefont {Szameit}},\ and\ \bibinfo {author}
  {\bibfnamefont {R.}~\bibnamefont {Thomale}},\ }\href@noop {} {\bibfield
  {journal} {\bibinfo  {journal} {Phys. Rev. Lett.}\ }\textbf {\bibinfo
  {volume} {126}},\ \bibinfo {pages} {215302} (\bibinfo {year}
  {2021})}\BibitemShut {NoStop}%
\bibitem [{\citenamefont {Chao}\ \emph {et~al.}(2024)\citenamefont {Chao},
  \citenamefont {Gokhale}, \citenamefont {Lin}, \citenamefont {Hastewell},
  \citenamefont {Bacanu}, \citenamefont {Chen}, \citenamefont {Li},
  \citenamefont {Liu}, \citenamefont {Lee}, \citenamefont {Dunkel},\ and\
  \citenamefont {Fakhri}}]{chao2024}%
  \BibitemOpen
  \bibfield  {author} {\bibinfo {author} {\bibfnamefont {Y.-C.}\ \bibnamefont
  {Chao}}, \bibinfo {author} {\bibfnamefont {S.}~\bibnamefont {Gokhale}},
  \bibinfo {author} {\bibfnamefont {L.}~\bibnamefont {Lin}}, \bibinfo {author}
  {\bibfnamefont {A.}~\bibnamefont {Hastewell}}, \bibinfo {author}
  {\bibfnamefont {A.}~\bibnamefont {Bacanu}}, \bibinfo {author} {\bibfnamefont
  {Y.}~\bibnamefont {Chen}}, \bibinfo {author} {\bibfnamefont {J.}~\bibnamefont
  {Li}}, \bibinfo {author} {\bibfnamefont {J.}~\bibnamefont {Liu}}, \bibinfo
  {author} {\bibfnamefont {H.}~\bibnamefont {Lee}}, \bibinfo {author}
  {\bibfnamefont {J.}~\bibnamefont {Dunkel}},\ and\ \bibinfo {author}
  {\bibfnamefont {N.}~\bibnamefont {Fakhri}},\ }\href
  {https://arxiv.org/abs/2410.18017} {\bibinfo {title} {Selective excitation of
  work-generating cycles in nonreciprocal living solids}} (\bibinfo {year}
  {2024}),\ \Eprint {https://arxiv.org/abs/2410.18017} {arXiv:2410.18017
  [cond-mat.soft]} \BibitemShut {NoStop}%
\bibitem [{\citenamefont {Fruchart}\ \emph
  {et~al.}(2023{\natexlab{b}})\citenamefont {Fruchart}, \citenamefont
  {Scheibner},\ and\ \citenamefont {Vitelli}}]{fruch23}%
  \BibitemOpen
  \bibfield  {author} {\bibinfo {author} {\bibfnamefont {M.}~\bibnamefont
  {Fruchart}}, \bibinfo {author} {\bibfnamefont {C.}~\bibnamefont
  {Scheibner}},\ and\ \bibinfo {author} {\bibfnamefont {V.}~\bibnamefont
  {Vitelli}},\ }\href@noop {} {\bibfield  {journal} {\bibinfo  {journal}
  {Annual Review of Condensed Matter Physics}\ }\textbf {\bibinfo {volume}
  {14}},\ \bibinfo {pages} {471} (\bibinfo {year}
  {2023}{\natexlab{b}})}\BibitemShut {NoStop}%
\bibitem [{\citenamefont {Ballantine}(1929)}]{Ballantine1929}%
  \BibitemOpen
  \bibfield  {author} {\bibinfo {author} {\bibfnamefont {S.}~\bibnamefont
  {Ballantine}},\ }\href {https://doi.org/10.1109/JRPROC.1929.221771}
  {\bibfield  {journal} {\bibinfo  {journal} {Proc. Inst. Radio Eng.}\ }\textbf
  {\bibinfo {volume} {17}},\ \bibinfo {pages} {927} (\bibinfo {year}
  {1929})}\BibitemShut {NoStop}%
\bibitem [{\citenamefont {Saadat}(1999)}]{saadat1999power}%
  \BibitemOpen
  \bibfield  {author} {\bibinfo {author} {\bibfnamefont {H.}~\bibnamefont
  {Saadat}},\ }\href {https://books.google.com/books?id=cVk2zgEACAAJ} {\emph
  {\bibinfo {title} {Power System Analysis}}}\ (\bibinfo  {publisher}
  {WCB/McGraw-Hill},\ \bibinfo {year} {1999})\BibitemShut {NoStop}%
\bibitem [{\citenamefont {Nagel}\ and\ \citenamefont {Pederson}(1973)}]{Spice}%
  \BibitemOpen
  \bibfield  {author} {\bibinfo {author} {\bibfnamefont {L.~W.}\ \bibnamefont
  {Nagel}}\ and\ \bibinfo {author} {\bibfnamefont {D.}~\bibnamefont
  {Pederson}},\ }\href
  {http://www2.eecs.berkeley.edu/Pubs/TechRpts/1973/22871.html} {\emph
  {\bibinfo {title} {SPICE (Simulation Program with Integrated Circuit
  Emphasis)}}},\ \bibinfo {type} {Tech. Rep.}\ \bibinfo {number} {UCB/ERL
  M382}\ (\bibinfo {year} {1973})\BibitemShut {NoStop}%
\bibitem [{LTs()}]{LTspice}%
  \BibitemOpen
  \href@noop {} {\bibinfo {title} {{LTspice 17.1}}},\ \bibinfo {note} {{Analog
  Devices Inc, Wilmington, MA, USA}}\BibitemShut {NoStop}%
\bibitem [{\citenamefont {Ashcroft}\ and\ \citenamefont
  {Mermin}(1976)}]{ashcroft1976solid}%
  \BibitemOpen
  \bibfield  {author} {\bibinfo {author} {\bibfnamefont {N.~W.}\ \bibnamefont
  {Ashcroft}}\ and\ \bibinfo {author} {\bibfnamefont {N.~D.}\ \bibnamefont
  {Mermin}},\ }\href@noop {} {\emph {\bibinfo {title} {Solid State Physics}}},\
  Vol.\ \bibinfo {volume} {2005}\ (\bibinfo  {publisher} {Saunders},\ \bibinfo
  {year} {1976})\BibitemShut {NoStop}%
\bibitem [{\citenamefont {Lin}\ \emph {et~al.}(2024)\citenamefont {Lin},
  \citenamefont {Yasuda}, \citenamefont {Ishimoto},\ and\ \citenamefont
  {Komura}}]{Lin2024}%
  \BibitemOpen
  \bibfield  {author} {\bibinfo {author} {\bibfnamefont {L.-S.}\ \bibnamefont
  {Lin}}, \bibinfo {author} {\bibfnamefont {K.}~\bibnamefont {Yasuda}},
  \bibinfo {author} {\bibfnamefont {K.}~\bibnamefont {Ishimoto}},\ and\
  \bibinfo {author} {\bibfnamefont {S.}~\bibnamefont {Komura}},\ }\href
  {https://doi.org/10.1103/PhysRevResearch.6.033016} {\bibfield  {journal}
  {\bibinfo  {journal} {Phys. Rev. Res.}\ }\textbf {\bibinfo {volume} {6}},\
  \bibinfo {pages} {033016} (\bibinfo {year} {2024})}\BibitemShut {NoStop}%
\bibitem [{\citenamefont {Nash}\ \emph {et~al.}(2015)\citenamefont {Nash},
  \citenamefont {Kleckner}, \citenamefont {Read}, \citenamefont {Vitelli},
  \citenamefont {Turner},\ and\ \citenamefont {Irvine}}]{nash2015topological}%
  \BibitemOpen
  \bibfield  {author} {\bibinfo {author} {\bibfnamefont {L.~M.}\ \bibnamefont
  {Nash}}, \bibinfo {author} {\bibfnamefont {D.}~\bibnamefont {Kleckner}},
  \bibinfo {author} {\bibfnamefont {A.}~\bibnamefont {Read}}, \bibinfo {author}
  {\bibfnamefont {V.}~\bibnamefont {Vitelli}}, \bibinfo {author} {\bibfnamefont
  {A.~M.}\ \bibnamefont {Turner}},\ and\ \bibinfo {author} {\bibfnamefont
  {W.~T.}\ \bibnamefont {Irvine}},\ }\href@noop {} {\bibfield  {journal}
  {\bibinfo  {journal} {Proc. Natl. Acad. Sci. U.S.A.}\ }\textbf {\bibinfo
  {volume} {112}},\ \bibinfo {pages} {14495} (\bibinfo {year}
  {2015})}\BibitemShut {NoStop}%
\bibitem [{\citenamefont {Wang}\ \emph {et~al.}(2021)\citenamefont {Wang},
  \citenamefont {Phan}, \citenamefont {Li}, \citenamefont {Wombacher},
  \citenamefont {Qu}, \citenamefont {Peng}, \citenamefont {Chen}, \citenamefont
  {Goldman}, \citenamefont {Levin}, \citenamefont {Austin},\ and\ \citenamefont
  {Liu}}]{wang2021emergent}%
  \BibitemOpen
  \bibfield  {author} {\bibinfo {author} {\bibfnamefont {G.}~\bibnamefont
  {Wang}}, \bibinfo {author} {\bibfnamefont {T.~V.}\ \bibnamefont {Phan}},
  \bibinfo {author} {\bibfnamefont {S.}~\bibnamefont {Li}}, \bibinfo {author}
  {\bibfnamefont {M.}~\bibnamefont {Wombacher}}, \bibinfo {author}
  {\bibfnamefont {J.}~\bibnamefont {Qu}}, \bibinfo {author} {\bibfnamefont
  {Y.}~\bibnamefont {Peng}}, \bibinfo {author} {\bibfnamefont {G.}~\bibnamefont
  {Chen}}, \bibinfo {author} {\bibfnamefont {D.~I.}\ \bibnamefont {Goldman}},
  \bibinfo {author} {\bibfnamefont {S.~A.}\ \bibnamefont {Levin}}, \bibinfo
  {author} {\bibfnamefont {R.~H.}\ \bibnamefont {Austin}},\ and\ \bibinfo
  {author} {\bibfnamefont {L.}~\bibnamefont {Liu}},\ }\href@noop {} {\bibfield
  {journal} {\bibinfo  {journal} {Phys. Rev. Lett.}\ }\textbf {\bibinfo
  {volume} {126}},\ \bibinfo {pages} {108002} (\bibinfo {year}
  {2021})}\BibitemShut {NoStop}%
\end{thebibliography}

%

\end{document}


\title{\Large Odd electrical circuits\\ \large Supplementary Information}

\author{Harry Walden}
\affiliation{Department of Mathematics, Massachusetts Institute of Technology, Cambridge, MA, USA}

\author{Alexander Stegmaier}
\affiliation{Institut f\"ur Theoretische Physik und Astrophysik, Universit\"at W\"urzburg, Würzburg, Germany}

\author{J\"orn Dunkel}
\affiliation{Department of Mathematics, Massachusetts Institute of Technology, Cambridge, MA, USA}

\author{Alexander Mietke}
\email{alexander.mietke@physics.ox.ac.uk}
\affiliation{Department of Physics, University of Oxford, Oxford, UK}

\date{\today}
\maketitle
\onecolumngrid
\tableofcontents

\begin{figure}[htpb]
    \centering
    \includegraphics[width=\linewidth]{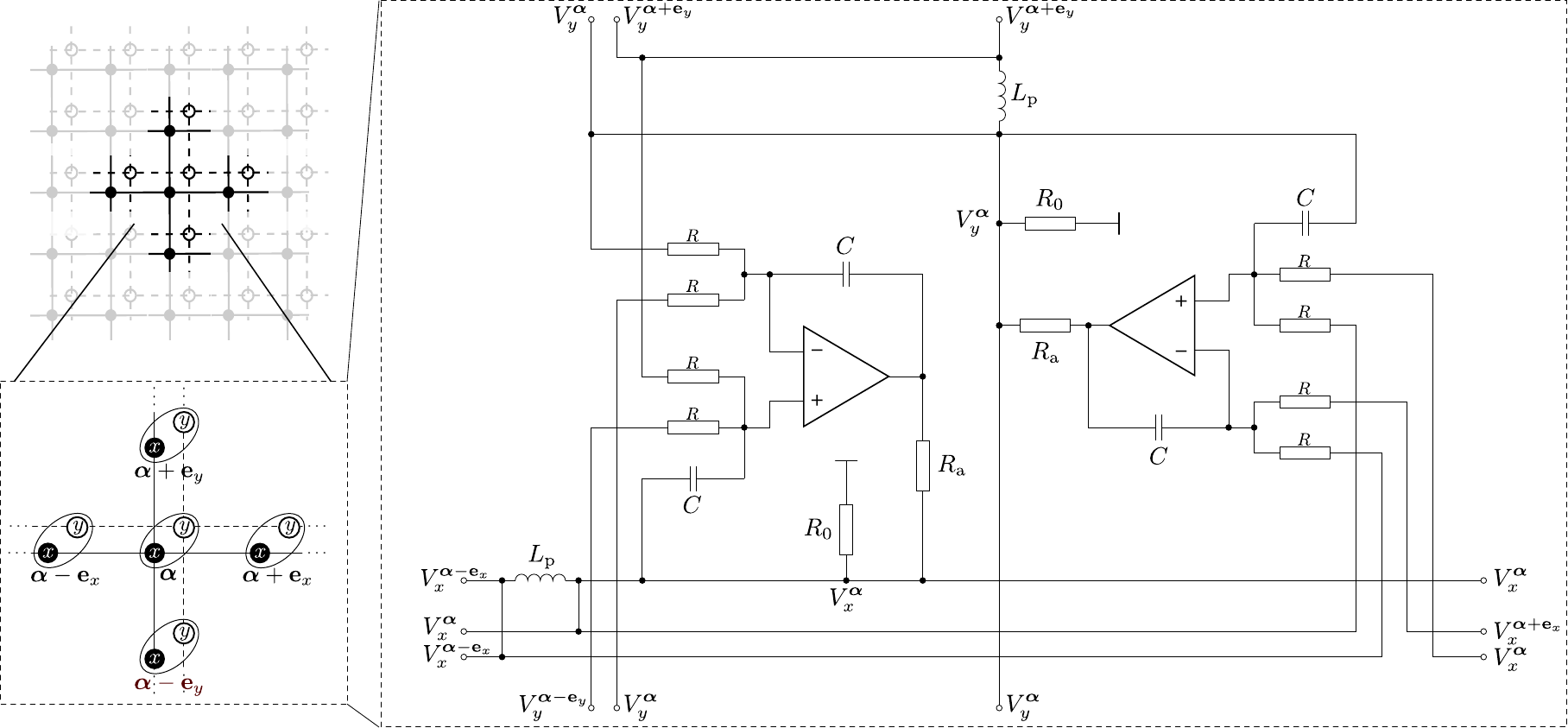}
    \caption{Circuit schematic of the odd elastic circuit's unit cell connecting lattice site $\bsa=(n,m)$ to lattice sites \hbox{$\bsa\pm\mathbf{e}_x=(n\pm1,m)$} and \hbox{$\bsa\pm\mathbf{e}_y=(n,m\pm1)$}. A lattice of arbitrary size can be constructed by connecting translated copies of the unit cell. The layout is configured so that with connectors at the boundary left open, the correct boundary conditions are implemented. Operational amplifiers in this circuit act as integrators and generate non-reciprocal current flows that emulate effectively negative inductances as discussed in main text. The supply voltage of operational amplifiers used to determine the internal power consumption $\mathcal{P}_{\text{oa}}$ in the main text, as well as serial resistances to inductors ($R_{L}$) and capacitors ($R_C$) that are part of realistic circuit element in LTspice simulations, are for brevity not shown in this circuit diagram. All parameter values used in all simulations are listed in Tab.~\ref{tab:tableS1}.}
    \label{fig:S1}
\end{figure}

\section{Electronic circuit design and properties}
In this section, we describe the implementation of an odd electric circuit (OEC) on a square lattice with nearest neighbor interactions. We then derive its current response and general spectral properties and prove the equivalence of this response with an odd spring network in the ideal circuit limit. 

\subsection{Unit cell design}
The circuit diagram of a complete unit cell of an odd electric circuit is shown in Fig.~\ref{fig:S1}. Nodes on $x-$ and $y-$sublattices are connected to their nearest neighbors in $x-$ and $y-$direction, respectively, by inductors with inductance $L_{\text{p}}$ to mediate the interactions analog to a passive spring (see Fig.~1d, main text). Non-reciprocal ``odd" couplings are implemented using active operational amplifier (OA) sub-circuits (see Fig.~1e, main text). The latter connect nodes $({\bsa},y)$ and $({\bsa}\pm\mathbf{e}_y,y)$ with $({\bsa},x)$, and nodes $({\bsa},x)$ and $({\bsa}\pm\mathbf{e}_x,x)$ with~$({\bsa},y)$ (left and right OA, respectively, in Fig.~\ref{fig:S1}). Output currents flow from the OAs through resistors~$R_{\text{a}}$ into nodes $({\bsa},x)$ and $({\bsa},y)$ and -- as shown below -- are proportional to differences in fluxes at corresponding $x$- and $y$-nodes as required by an odd spring interaction. The resistance $R$ of resistors within the OA subcircuits are chosen sufficiently large, so that spurious currents between, for example, the directly connected nodes $({\bsa}+\mathbf{e}_x,x)$ and $({\bsa}-\mathbf{e}_x,x)$ are negligible.

\subsection{Reciprocity and mutuality in electrical circuits}

As indicated in the main text, we follow the convention from~\cite{braverman2020topological} and distinguish between \textit{reciprocity} related to the work performed by a system and discussed in more detail below, and \textit{mutuality}, describing forces (or currents, under the mobility analogy) which are equal and opposite among interacting particles or nodes. 

Odd elastic springs describe non-reciprocal interactions in the sense that they violate the Maxwell-Betti work theorem~\cite{Scheibner:2020gm,braverman2020topological}, which is a theorem that must hold in conservative systems. In short, consider two sets of forces $F^P_i$ and $F^Q_i$ are applied at locations $i=1,...,n$ and determine the resulting (linear response) displacements $u_i^P$ and $u_i^Q$ at the same locations, then Maxwell-Betti reciprocity holds when the relation
\begin{equation}
    \sum_{i=1}^n F_i^P u_i^Q = \sum_{i=1}^n F_i^Q u_i^P
    \label{eq:maxwell-betti}
\end{equation}
is satisfied. Without further calculation, the equivalent result for a reciprocal circuit in the mobility analogy must therefore satisfy
\begin{equation}
    \sum_{i=1}^n I_i^P \phi_i^Q = \sum_{i=1}^n I_i^Q \phi_i^P,
    \label{eq:mb electrical}
\end{equation}
which we can identify as a special case of Lorentz reciprocity in circuits \cite{Ballantine1929}. Since the mechanical system is known to violate Maxwell-Betti reciprocity, it must also be the case that the electrical analog violates Lorentz reciprocity. 

\subsection{Derivation of current-flux response in a realistic circuit}
We now derive the form of the current at some lattice site $\bsa$ for the circuit unit cell shown in Fig.~\ref{fig:S1}. To compute active currents, we take into account a realistic OA dynamics and show that, in the limit of ideal circuit elements (no variability of circuit element properties, no parasitic resistances and an OA with infinite gain) the currents generated at each lattice side provide a one-to-one analog with an odd spring network, effectively emulating the inverse inductance matrix $\Lambda_{ij}^{\bsa\bsb}$ introduced in the main text. Specifically, we will show that that the currents at a lattice site $\bsa$ obey for ideal circuit elements
\begin{align}
I_x^{\bsa}&=\frac{1}{L_{\text{p}}}\left(\phi_x^{\bsa+\mathbf{e}_x}+\phi_x^{\bsa-\mathbf{e}_x}-2\phi_x^{\bsa}\right)+\frac{1}{R_{\text{a}}RC}\left(\phi_y^{\bsa+\mathbf{e}_y}+\phi_y^{\bsa-\mathbf{e}_y}-2\phi_y^{\bsa}\right)\label{eq:SiIx}-\frac{1}{R_0}V_x^{\bsa}\\
I_y^{\bsa}&=\frac{1}{L_{\text{p}}}\left(\phi_y^{\bsa+\mathbf{e}_y}+\phi_y^{\bsa-\mathbf{e}_y}-2\phi_y^{\bsa}\right)-\frac{1}{R_{\text{a}}RC}\left(\phi_x^{\bsa+\mathbf{e}_x}+\phi_x^{\bsa-\mathbf{e}_x}-2\phi_x^{\bsa}\right)-\frac{1}{R_0}V_y^{\bsa}.\label{eq:SiIy}
\end{align}
The first contribution on the right-hand side corresponds to a classic passive spring response with ``spring constant" $k_{\text{p}}:=L_{\text{p}}^{-1}$. The second contribution is proportional to an effective inverse inductance $k_{\text{a}}:=(R_{\text{a}}RC)^{-1}$, which represents the analog of an odd spring constant. Identifying currents with forces and fluxes with displacements (mobility analogy), these terms represent the electronic analog of an odd elastic spring (compare with Eq.~(2), main text). In addition, Eqs.~(\ref{eq:SiIx}) and~(\ref{eq:SiIy}) contain contributions $\sim R_0^{-1}$, the electrical analog of mechanical friction.

To show that the circuit in Fig.~(\ref{fig:S1}) is described by Eqs.~(\ref{eq:SiIx}) and~(\ref{eq:SiIy}), we will interchangeably use real-space and plane-wave representations of voltage modes at a given lattice node $\bsa=(n,m)$ of the $x$- or $y$ sub-lattice, i.e.
\begin{equation}\label{eq:PW}
V^{\bsa}_i(t)=v_i\exp[i(q_xn+q_ym)a]\exp(i\omega t),
\end{equation}
where $i=x,y$ indicates the sub-lattice. 

\subsubsection{Passive subcircuit}
In the passive sub-circuit, nearest neighbor nodes are coupled via passive inductors (see Fig.~\ref{fig:S1}) and each node is grounded via a resistor $R_0$, which directly gives rise to the terms $\sim L^{-1}_{\text{p}}$ and $\sim R_0^{-1}$ in Eqs.~(\ref{eq:SiIx}),(\ref{eq:SiIy}). In Fourier-space, these contributions take the form
\begin{align}\label{eq:ip}
\begin{pmatrix}
i_x \\ i_y
\end{pmatrix}&=
\frac{2}{i\omega L_{\text{p}}}
\begin{pmatrix}
\cos q_x -1 & 0 \\ 0 & \cos q_y-1
\end{pmatrix}
\begin{pmatrix}
v_x \\ v_y
\end{pmatrix}-\frac{1}{R_0}
\begin{pmatrix}
1& 0 \\ 0 & 1
\end{pmatrix}
\begin{pmatrix}
v_x \\ v_y
\end{pmatrix}.
\end{align}

\subsubsection{Active subcircuit}
The output voltage of OAs can be computed as follows. First, we solve for the voltages $V^-$, $V^+$ at the OA's $``-"$ and $``+"$ inputs, and the voltage $V^o$ at its output. Note that OA-related voltages $V^-_i$, $V^+_i$ and $V^o_i$ are always local to a single lattice site and will be written as $V_i(t)=v_ie^{i\omega t}$ with corresponding superscripts $-,+,o$ to be added.\\

\textbf{Active currents in $\boldsymbol{x}$-direction: } Active currents send by OAs to neighbor nodes in $x$-direction (OA on the left in Fig.~\ref{fig:S1}) can be determined from Kirchhoff's law as follows.
\begin{itemize}
\item The currents at the OP-amp $\lq-\rq$ input node have to add to zero, which yields 
\begin{align}
0&=C\frac{d}{dt}(V^o_x-V^-_x)+\frac{2}{R}(V^{\bsa}_y-V^-_x)
.\label{eq:cb1}
\end{align}
The first term is the current across the feedback capacitor connecting the OA output $V_o$ to its $``-"$ input. The second term describes the current between voltage node $V^{\bsa}_y$ and the $``-"$ input of the OA. The current balance Eq.~(\ref{eq:cb1}) can be solved in Fourier space for the voltage at the $``-"$ input of the OA, where we find
\begin{align}
v^-_x&=\frac{i\omega\tau}{i\omega\tau+2}v^o_x+\frac{2}{i\omega\tau+2}v_y,\label{eq:Vxminus}
\end{align}
where we introduced for convenience the time scale 
\begin{equation}
\tau=RC.    
\end{equation}

\item Currents at the OA $``+"$ input node have to add to zero as well, such that Kirchhoff's law implies
\begin{align}
&=C\frac{d}{dt}(V_x^{\bsa}-V^+_x)+\frac{1}{R}\left(V^{\bsa+\mathbf{e}_y}_y-V^+_x\right)+\frac{1}{R}\left(V^{\bsa-\mathbf{e}_y}-V^+_x\right)
\nonumber.
\end{align}
The first term is the current flowing from the $V_{\bsa}^x$ node through the capacitor with capacity $C$ into the OA's $``+"$ input node. The other terms contribute in a similar fashion as for the computation of the current balance at the OAs $``-"$ input. Using the plane wave ansatz, this can be solved for the voltage at the $``+"$ input of the OA
\begin{align}
v^+_x&=\frac{i\tau\omega}{i\tau\omega+2}v_x+\frac{2\cos q_y}{i \tau\omega+2}v_y.\label{eq:Vxplus}
\end{align}
\item Finally, we require a gain model for the OA. A simple, realistic representation of an OA response is given in terms of an active low-path filter coupling input voltage difference $V^+_x-V^-_x$ to an output voltage~$V^o_x$ via
\begin{equation}\label{eq:opamp}
\left(1+\omega_0^{-1}\frac{d}{dt}\right)V^o_x=G(V^+_x-V^-_x)\Rightarrow v^o_x=\frac{G}{1+i\omega/\omega_0}(v^+_x-v^-_x),
\end{equation}
where $\omega_0=C_{amp}R_{amp}$ is the low-pass cut-off frequency and $G$ is the gain factor. For any lattice site voltage modes $(v_x,v_y)$, Eqs.~(\ref{eq:Vxminus}), (\ref{eq:Vxplus}) and (\ref{eq:opamp}) provide a system of equations that can be solved for the voltages $v^-_x$, $v^+_x$ and $v^o_x$ at the OA's input and output. Of particular interest is the OA's output voltage mode~$v^o_x$, which sets the current flowing into the local $x$-node and is given by
\begin{equation}
v^o_x=\frac{i\omega\tau+2}{(1+i\omega/\omega_0)(i\omega\tau+2)/G+i\omega\tau}\left[\frac{i\tau\omega}{i\tau\omega+2}v_x+\frac{2\cos q_y-2}{i\tau\omega+2}\right].\label{eq:vox}
\end{equation}

Finally, the active current flowing into the $x$-node is in mode space given by
\begin{equation}\label{eq:ixop}
I^{\bsa}_x=\frac{V^o_x-V_x^{\bsa}}{R_\text{a}}\Leftrightarrow i_x=\frac{v_x^o-v_x}{R_\text{a}}.
\end{equation}
Note, in the limit of infinite gain $G\rightarrow\infty$ (ideal OA), we have
\begin{equation}
v^o_x\rightarrow\frac{1}{i\omega RC}(2\cos q_y-2)v_y+v_x
\end{equation}
and therefore the current simplifies to
\begin{equation}
i_x=\frac{1}{i\omega R_\text{a}RC}(2\cos q_y-2)v_y\Leftrightarrow I^{\bsa}_x=\frac{1}{R_\text{a}RC}(\phi_y^{\bsa-\mathbf{e}_y}+\phi_y^{\bsa+\mathbf{e}_y}-2\phi_y^{\bsa}).
\end{equation}
The last expression corresponds to the active currents in real space. Indeed, this expression agrees with the active contribution in Eq.~(\ref{eq:SiIx}) and analog to neighbor-site interactions via odd springs.
\end{itemize}

\noindent\textbf{Active currents in $\boldsymbol{y}$-direction: }A similar calculation determines the currents send by OAs to neighboring nodes in the $y$-direction (OA on the right in Fig.~\ref{fig:S1}).
\begin{itemize}
\item From current balance at "$-$" input of the OA, we find
\begin{align*}
0&=C\frac{d}{dt}(V_y^o-V^-_y)+\frac{1}{R}\left(V^{\bsa+\mathbf{e}_x}_x-V^-_y\right)+\frac{1}{R}\left(V^{\bsa-\mathbf{e}_x}_x-V^-_y\right)
\nonumber.
\end{align*}
This yields in mode space
\begin{align}
v^-_y&=\frac{i\omega\tau}{i\omega\tau+2}v^o_x+\frac{2\cos q_x}{i\omega\tau+2}v_x.\label{eq:Vyminus}
\end{align}
\item From current balance at "$+$"-OA input, we find
\begin{align*}
0&=C\frac{d}{dt}(V^o_y-V^+_y)+\frac{2}{R}(V^{\bsa}_x-V^+_y),
\end{align*}
which yields in mode space
\begin{align}
v^+_y&=\frac{i\tau\omega}{i\tau\omega+2}v_y+\frac{2}{i\tau\omega+2}v_x.\label{eq:Vyplus}
\end{align}
\item The OA model is the same as before. Specifically, for the OA on the right in Fig.~\ref{fig:S1}, we have
\begin{equation}\label{eq:opamp_rep}
\left(1+\omega_0^{-1}\frac{d}{dt}\right)V_o^y=G(V_+^y-V_-^y)\Rightarrow v^o_y=\frac{G}{1+i\omega/\omega_0}(v^+_y-v^-_y),
\end{equation}
such that we find from Eqs.~(\ref{eq:Vyminus}), (\ref{eq:Vyplus}) and (\ref{eq:opamp_rep}) the OA's output voltage
\begin{equation}\label{eq:voy}
v^o_y=\frac{i\omega\tau+2}{(1+i\omega/\omega_0)(i\omega\tau+2)/G+i\omega\tau}\left[\frac{2-2\cos q_x}{i\tau\omega+2}v_x+\frac{i\tau\omega}{i\tau\omega+2}v_y\right].
\end{equation}
The overall output current is given by
\begin{equation}\label{eq:iyop}
I_y^{\bsa}=\frac{V^o_y-V^{\bsa}_y}{R_{\text{a}}}\Leftrightarrow i_y=\frac{v^o_y-v_y}{R_{\text{a}}}
\end{equation}
and in the limit of an ideal OA and vanishing capacitor asymmetry we find
\begin{equation}
v^o_y\rightarrow\frac{1}{i\omega RC}(2-2\cos q_x)v_x+v_y
\end{equation}
and therefore the current simplifies to
\begin{equation}
i_y=\frac{1}{i\omega R_{\text{a}}RC}(2-2\cos q_x)v_x\Leftrightarrow I^{\bsa}_y=-\frac{1}{R_{\text{a}}RC}(\phi^{\bsa-\mathbf{e}_x}_x+\phi^{\bsa+\mathbf{e}_x}_x-2\phi^{\bsa}_x).
\end{equation}
This confirms the active contributions in Eq.~(\ref{eq:SiIy}), which are analog to neighbor-site interactions via odd springs.
\end{itemize}

\subsection{Dispersion relation}
We can now write down a full spectral description of the dynamics odd electric circuit model. From Eqs.~(\ref{eq:ip}), (\ref{eq:ixop}) and (\ref{eq:iyop}), we find that the circuit's current response is described by
\begin{align}
\begin{pmatrix}
i_x \\ i_y
\end{pmatrix}&=\mathcal{H}(q_x,q_y,\omega)\begin{pmatrix}
v_x \\ v_y
\end{pmatrix}
\nonumber\\
&:=\left[\frac{1}{R_{\text{a}}}
\begin{pmatrix}
[v_x^o]_{x} - 1& [v_x^o]_{y}\\
[v_y^o]_{x} & [v_y^o]_{y} - 1 &
\end{pmatrix}+
\frac{2}{i\omega L_\text{p}}
\begin{pmatrix}
\cos q_x-1 & 0 \\ 0 & \cos q_y-1
\end{pmatrix}-
\frac{1}{R_0}\begin{pmatrix}
1 & 0 \\ 0 & 1
\end{pmatrix}\right]
\begin{pmatrix}
v_x \\ v_y
\end{pmatrix},\label{eq:idyn}
\end{align}
where $[v_x^o]_{x}$ and $[v_y^o]_{y}$ are given by
\begin{align}
[v_x^o]_{x}&=[v_y^o]_{y}=\frac{i\tau\omega}{(1+i\omega/\omega_0)(i\omega\tau+2)/G+i\omega\tau}\\
[v_x^o]_{y}&=\frac{2\cos q_y-2}{(1+i\omega/\omega_0)(i\omega\tau+2)/G+i\omega\tau}\\
[v_y^o]_{x}&=\frac{2-2\cos q_x}{(1+i\omega/\omega_0)(i\omega\tau+2)/G+i\omega\tau},
\end{align}
as implicitly defined in Eqs.~(\ref{eq:vox}) and (\ref{eq:voy}). The self-organized response of the circuit follows from Kirchhoff's law as
\begin{equation}\label{eq:KH}
\begin{pmatrix}
i_x \\ i_y
\end{pmatrix}=0.
\end{equation}
Accordingly, $\det{\mathcal{H}}=0$ with $\mathcal{H}(q_x,q_y,\omega)$ defined in Eq.~(\ref{eq:idyn}) determines the circuit's dispersion relation~$\omega(q_x,q_y)$. In the limit of an ideal circuit, in particular an OA with infinite gain $G\rightarrow\infty$, we find from Eq.~(\ref{eq:idyn})
\begin{equation}
\mathcal{H}=
\begin{pmatrix}
\frac{2k_{\text{p}}}{i\omega}(\cos q_x-1)-\sigma & \frac{2k_{\text{a}}}{i\omega}(\cos q_y-1)\\
\frac{2k_{\text{a}}}{i\omega}(1 - \cos q_x) & \frac{2k_{\text{p}}}{i\omega}(\cos q_y-1)-\sigma
\end{pmatrix},
\end{equation}
where $k_{\text{p}}=L_{\text{p}}^{-1}$, $k_\text{a}=(R_{\text{a}}RC)^{-1}$ and $\sigma=1/R_0$. Thus, for $\sigma>0$, we have
\begin{align}
i\sigma\omega&=-2k_{\text{p}}\left(\sin^2\frac{q_x}{2} + \sin^2\frac{q_y}{2}\right)\pm2\sqrt{k_{\text{p}}^2\sin^2\frac{q_x-q_y}{2}\sin^2\frac{q_x+q_y}{2}-4k_{\text{a}}^2\sin^2\frac{q_x}{2}\sin^2\frac{q_y}{2}}.
\end{align}
For diagonal wave vectors of the form $\mathbf{q}=(q,q)/\sqrt{2}$, the dispersion relation simplifies to
\begin{align} \label{seq:disp rel}
i\sigma\omega&=-4\left(k_{\text{p}}\pm i|k_{\text{a}}|\right)\sin^2\left(\frac{q}{2\sqrt{2}}\right).
\end{align}

\section{Mechanical-electrical analogy}
\begin{table}[h!]
\centering
\begin{tabular}{c|c|c}
& Mobility analogy & Impedance analogy\\ 
\specialrule{.1em}{.2em}{0.2em}
Force & Current~$\mathbf{I}$ & Voltage~$\mathbf{V}$ \\\hline 
Displacement & Flux~$\boldsymbol{\phi}=\int\mathbf{V}dt$ & Charge~$\mathbf{Q}=\int\mathbf{I}dt$ \\\hline
Work & $\mathbf{I}\cdot d\boldsymbol{\phi}=\mathbf{I}\cdot\mathbf{V}dt$ & $\mathbf{V}\cdot d\boldsymbol{Q}=\mathbf{V}\cdot\mathbf{I}dt$
\end{tabular}
\caption{Key identifications in mechanical-electrical analogies. The present work employs the mobility analogy.}
\label{tab:analogs}
\end{table}

As explained in the main text, the mobility analogy is distinct as it allows for a direct identification of mechanical and electrical work, i.e. $dW_{\text{mech}}=\mathbf{F}\cdot d\mathbf{u}\Leftrightarrow dW_{\text{el}}=\mathbf{I}\cdot d\boldsymbol{\phi}=\mathbf{I}\cdot\mathbf{V}dt$. While this identification is also possible using an impedance analogy (see Tab.~\ref{tab:analogs}), we explain in the following why the latter is not a suitable choice for studying the electrical analog of a force-response to imposed deformations on the nearest-neighbor square-lattice topology considered in this work.  

\subsection{Mobility analogy}
We start by rewriting for convenience Eq.~(13) of the main text in the form
\begin{equation}\label{eq:KLsi}
\mathbf{I}=-\mathbf{\Lambda}\cdot\boldsymbol{\phi}-\sigma\frac{d\boldsymbol{\phi}}{dt},
\end{equation}
where $\mathbf{\Lambda}$ is the coupling matrix containing inverse inductances. For any spatial distribution of displacement-analog fluxes $\boldsymbol{\phi}$, Eq.~(\ref{eq:KLsi}) yields a unique force-analog current response, as required. 

In anticipation of explaining the shortcomings of the impedance analogy below, we note the following. If one would be interested instead in a flux (displacement) response for given currents (forces), Eq.~(\ref{eq:KLsi}) has to be inverted. This requires that $\mathbf{\Lambda}$ is invertible, which is not the case here: For nearest-neighbor interactions on a square lattice that was chosen in this work for simplicity, flux patterns $\Phi^1$ and $\Phi^3$ for example do not lead to currents (see Eq.~(12), main text, and following discussion). Consequently, $\mathbf{\Lambda}$ is not invertible on a square lattice with nearest neighbor coupling. 

\subsection{Impedance analogy}
An alternative well-known mechanical-electrical analogy is the \textit{impedance analogy}, in which voltages $\mathbf{V}$ are identified with forces and charges $\mathbf{Q}$ with displacements. In this analogy, inverse capacitances take the role of mechanical spring constants. For a coupling matrix $\mathbf{C}$ containing inverse capacitances, we have in this case schematically a circuit dynamic described by
\begin{equation}\label{eq:V1}
\mathbf{V}=-\mathbf{C}\cdot\mathbf{Q}\Rightarrow\frac{d\mathbf{V}}{dt}=-\mathbf{C}\cdot\mathbf{I}
\end{equation}
for interactions between nodes, and 
\begin{equation}\label{eq:V2}
\mathbf{V}=-R_0\frac{d\mathbf{Q}}{dt}=-R_0\mathbf{I}
\end{equation}
implementing local ``friction" by connecting each node additionally to ground. The collective response of the circuit is still determined by Kirchhoff's junction rule for the sum of currents at each node, which implies for Eqs.~(\ref{eq:V1}) and (\ref{eq:V2}) that currents at each node are given~by
\begin{equation}\label{eq:QV}
\mathbf{I}=-\mathbf{C}^{-1}\cdot\frac{d\mathbf{V}}{dt}-\sigma\mathbf{V}\Rightarrow\mathbf{Q}=-\mathbf{C}^{-1}\mathbf{V}-\sigma\int\mathbf{V}dt.
\end{equation}
The last expression, in principle, describes a convenient circuit analog to study collective displacements (represented by charges $\mathbf{Q}$) in response to imposed forces (represented by voltages $\mathbf{V}$), as opposed to an analog force response under imposed displacements (see Eq.~(\ref{eq:KLsi}) and Tab.~\ref{tab:analogs}). Equation~(\ref{eq:QV}) can only be evaluated for given voltages when $\mathbf{C}$ is invertible. In fact, even for imposed charges or currents, Eq.~(\ref{eq:QV}) cannot be solved uniquely for the resulting voltages when~$\mathbf{C}$ is not invertible and $\sigma\ne0$. However, on a square lattice with nearest neighbor interactions, the coupling matrix $\mathbf{C}$ will be proportional to~$\mathbf{\Lambda}$ and thus is not invertible. This renders the impedance analogy ill-defined for the minimal example of a nearest-neighbor square lattice topology considered in this work.\\

Finally, we note that $\mathbf{\Lambda}$ and Eq.~(\ref{eq:KLsi}), and consequently $\mathbf{C}$ and Eq.~(\ref{eq:QV}), all become invertible if, for example,  next-to-nearest (diagonal) neighbor couplings are added to the square-lattice circuit or if the topology of a hexagonal lattice with nearest-neighbor couplings is considered. In this case, both analogies in Tab.~\ref{tab:analogs} can in principle be used to implement circuits in which force/deformation-analog measurements in response to corresponding deformation/force-analogs can be performed.

\subsection{Continuum limit}
Using standard techniques~\cite{ashcroft1976solid}, one can show that spring interactions according to Eq.~(2) in the main text lead in the long-wavelength limit to a linearly elastic constitutive law $\sigma^{\mu}=C^{\mu\nu}u^{\nu}$, where $u^{\nu}=\partial_iu_j\tau^{\nu}_{ij}$ are the strain tensor components expressed in the $\tau$-tensor basis [see main text Eq.~(11)] and
\begin{equation}
C^{\mu\nu}=
\begin{pmatrix}
k_{\text{p}} & 0 & 0 & 0\\
k_{\text{a}} & 0 & 0 & 0 \\
0 & 0 & k_{\text{p}} & 0 \\
0 & 0 & -k_{\text{a}} & 0 
\end{pmatrix}\label{eq:modtens}
\end{equation}
is the elastic modulus tensor. For linear displacement gradients -- as considered in the main text for cyclic excitations -- the analog ``microscopic" constitutive law Eq.~(10) for OECs is therefore equivalent to the mean-field result for an odd elastic spring network on square lattice. Cartesian components of the stress, $\sigma_{ij}=\frac{1}{2}\sigma^{\mu}\tau^{\mu}_{ij}=\frac{1}{2}C^{\mu\nu}u^{\nu}\tau^{\mu}_{ij}$, are explicitly
\begin{equation}
\boldsymbol{\sigma}=
\begin{pmatrix}
k_{\text{p}}\partial_xu_x& -k_{\text{a}}\partial_xu_x\\
k_{\text{a}}\partial_yu_y & k_{\text{p}}\partial_yu_y 
\end{pmatrix},
\end{equation}
and the force balance is accordingly given by
\begin{equation}
\nabla\cdot\boldsymbol{\sigma}=
k_{\text{p}}
\begin{pmatrix}
\partial^2_xu_x\\
\partial^2_yu_y
\end{pmatrix}+
k_{\text{a}}
\begin{pmatrix}
\partial_y^2u_y\\
-\partial_x^2u_x
\end{pmatrix}.\label{eq:FBCG}
\end{equation}
The shear contributions $\mu=2,3$ in the modulus tensor Eq.~(\ref{eq:modtens}) indicate a manifestly anisotropic material~\cite{Scheibner:2020gm} even in the long-wavelength limit. This is a consequence of starting from a square lattice with nearest-neighbor interactions and different for hexagonal spring lattices, that coarse-grain into fully isotropic constitutive laws. Due to the lack of rotational invariance encountered in our system, force-balance equations (\ref{eq:FBCG}) cannot be immediately written in terms of coordinate free expressions.


\section{\emph{In silico} experimental setup}
Most of the numerical results in this work have been obtained using the circuit simulation software LTSpice~\cite{LTspice}, a freely available SPICE (Simulation Program with Integrated Circuit Emphasis) simulator that allows simulating complex networks of realistic electronic circuit elements. An overview of the circuit element parameters for a circuit with unit cell shown in Fig.~\ref{fig:S1} is provided in Tab.~\ref{tab:tableS1}. In the following, we describe these simulations in more detail.

\subsection{LTSpice simulations}

\begin{table}[h]
    \centering
    \begin{tabular}{c|c|c|c|}
    \vspace*{-5mm}
    \hspace*{7mm}&\hspace*{2cm}&\hspace*{2cm}&\hspace*{2cm}\\
         &  \parbox[b]{2cm}{Static \\ \vspace*{-1mm}response} & \parbox[b]{2cm}{Power\\ \vspace*{-1mm}extraction} & \parbox[b]{2cm}{Active\\ \vspace*{-1mm}resonances}\\
        \hline
        $R_0$ & --- & \multicolumn{2}{c|}{$\SI{10}{\ohm}$}\\ \hline
        $L_{\text{p}}$ &  \multicolumn{3}{c|}{\SI{0.1}{\milli\henry}}\\\hline
        $R_{L_\text{p}}$ & \multicolumn{3}{c|}{\SI{0.7}{\ohm}}\\\hline
        $R_{\text{a}}$ & \SI{100}{\ohm} &\multicolumn{2}{c|}{\SI{10}{\ohm}}\\\hline
        $C$ & \multicolumn{3}{c|}{\SI{100}{\pico\farad}}\\\hline
        $R_{C}$ & \multicolumn{3}{c|}{\SI{0.3}{\ohm}}\\\hline
        $R$ & \multicolumn{3}{c|}{\SI{10}{\kilo\ohm}}\\\hline
        $\omega_0$ & \multicolumn{3}{c|}{\SI{100}{\kilo\hertz}}\\\hline
        $\tilde{k}_{\text{p}}$ & $1$ & $1$ &$1$\\\hline
        $\tilde{k}_\text{a}$ & $1$ &$10$ &$10$\\\hline
    \end{tabular}
    \caption{Parameter values used in LTspice simulations of a circuit with unit cell elements shown in Fig.~\ref{fig:S1}. The table additionally includes realistic values of parasitic serial resistances of the passive inductors ($R_{L_\text{p}}$) and of the capacitor ($R_{C}$), which are part of any realistic circuit element and used accordingly in LTSpice simulations. For convenience, we introduce an arbitrary reference frequency~$\omega_0$ and denote dimensionless electrical analogs of passive and active (odd) springs by $\tilde{k}_{\text{p}}=R_0/(\omega_0L_\text{p})$ and \hbox{$\tilde{k}_{\text{a}}=R_0/(\omega_0CRR_{\text{a}})$}, respectively.}
    \label{tab:tableS1}
\end{table}

\subsubsection{Static spatial flux patterns}

Numerical simulations were carried out on $N\times N$ lattices by joining copies of the unit cell design Fig.~\ref{fig:S1}, leaving boundaries open. Such open connections draw can therefore draw currents correspond in the mechanical picture to stress-free boundaries.

Static tests were carried out for $N=5$ to verify the instantaneous current response to spatial flux gradients. This response was derived in the main text to be of the form
\begin{equation}\label{eq:J Phi coupling SI}
\begin{pmatrix}
\mathcal{I}^0\\
\mathcal{I}^1
\end{pmatrix}=
\begin{pmatrix}
k_{\text{p}}\\
k_{\text{a}} 
\end{pmatrix}\Phi^0
\quad\text{and}\quad
\begin{pmatrix}
\mathcal{I}^2\\
\mathcal{I}^3    
\end{pmatrix}=
\begin{pmatrix}
k_{\text{p}}\\
-k_{\text{a}}
\end{pmatrix}
\Phi^2.
\end{equation}

Simulations were set up using variable voltage sources across each node, and carried out in the time domain in LTSpice. The desired flux profile was initialized in each case by applying a voltage pulse of the appropriate magnitude and uniform width at each node. Circuit responses were determined from simulations immediately after application of the pulse. Ideal circuit theory predicts an instantaneous current response only as long as fluxes are present, but realistic losses in LTspice circuit simulations lead to a response that decays on a timescale much longer than that of the pulse.

\begin{figure}
    \centering
    \includegraphics[width=0.7\linewidth]{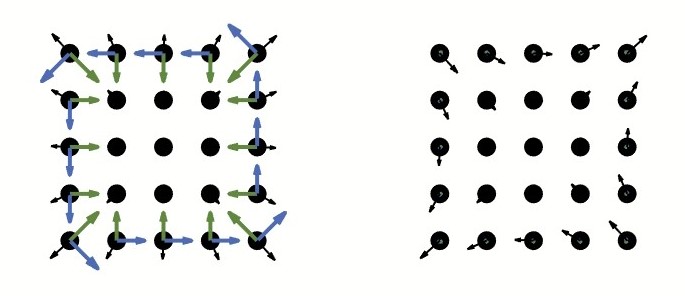}
    \includegraphics[width=0.7\linewidth]{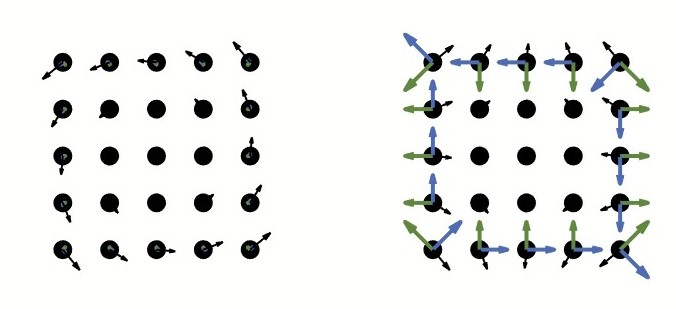}
    \caption{Visualization of driving fluxes and net corresponding net currents. Black arrows indicate vectorized fluxes, green arrows indicate vectorized passive currents and  blue arrows indicate vectorized active currents. Patterns correspond to isotropic expansion (top-left), rotation (bottom-left), shear 2 (top-right) and shear 1 (bottom-right). In accordance with Eq.~(12), only compression and shear 1 fluxes lead to non-zero components of the incident current tensor $\mathcal I$, which then cancel to give no net current at bulk sites.}
    \label{sfig: static patterns}
\end{figure}

As predicted by Eq.~(\ref{eq:J Phi coupling SI}), there was a negligible current response to rotation-like and diagonal shear-like flux gradients $\Phi^1$ or $\Phi^3$, respectively. For isotropic dilation-like and transverse shear-like flux gradients $\Phi^0$ and $\Phi^2$, respectively large currents were drawn. These large currents appear at boundary sites of the lattice where they support internal current flows that are expected to cancel each other in the bulk of the lattice. Indeed negligible current was drawn from voltage sources at interior lattice sites, which is consistent with the prediction in Eq.~(\ref{eq:J Phi coupling SI}) derived in the main text: An incident current tensor $\mathcal{I}$ that is non-zero but spatially constant corresponds to a case of of vanishing net currents in the bulk, where all incident currents are balanced by outgoing currents. At boundary sites, the inability to draw current from open connections requires drawing current from voltage sources to compensate~(Fig.~\ref{sfig: static patterns}). This is analog to elastostatics, where constant stresses imply a vanishing force-density in the bulk of the material, but lead to local net forces (in our case currents) at the boundary~\cite{landau1986}. These results deviate within a tolerance of 5\% from the ideal circuit theory in the specified parameter range~(Table~\ref{tab:tableS1}, \textit{Static response}). We attribute these small variations to parasitic resistances and variations in component parameters under the prescribed realistic tolerances that are implemented in LTSpice simulations. 

\subsubsection{Apparent power extraction cycles}\label{ssec: ltspice work extraction}

Numerical experiments to verify power extraction properties of OECs were similarly carried out in the time domain on a $5 \times 5$ lattice, and nodes were connected to a prescribed, time-varying voltage (i.e. flux) pattern. Here, the voltage pulse was replaced by oscillatory signals with slowly varying frequency in an effective AC sweep. Another notable difference from static tests is the inclusion of damping elements with conductance $\sigma=1/R_0$ as shown in the complete unit cell diagram (Fig.~\ref{fig:S1}). The full list of parameter values used for circuit elements in these simulations is provided in Table~\ref{tab:tableS1}~(\textit{Power extraction}).

Data in Figs.~2b,c of the main text were extracted from the time series data generated by LTSpice simulations. In particular, Fig.~2b was produced by extracting flux, current and power values from several periods around the theoretical peak output $\omega^*$. The sample frequency is much greater than the oscillation frequency, so the output appears as a continuous signal. Discrete points in Fig.~2c (main text) were determined by computing local averages with a bucketing procedure. The underlaid continuous lines are fits to the data based on the ideal circuit prediction Eq.~(19) in the main text. Indeed, such a fit recovers the microscopic circuit properties, the active spring constant analog $k_{\text{a}}=1/(CRR_{\text{a}})$ and the friction-analog conductance $\sigma$, within 5\% of their true input values. 

\subsubsection{Amplification and active resonance regimes}
To study amplification and resonance properties of OECs, we first considered a minimal example on a $3\times 3$ lattice and a piston-analog setup on a $15 \times 15$ lattice in the second step (See Fig.~3, main text).

A schematic representation of the minimal example lattice is shown in the main text (Fig.3a). To carry out the \textit{in-silico }experiments, an additional copy of the unit cell was appended to the left edge (-ve $x$-direction), which had its $x$-node voltage prescribed by an AC source. The top and bottom row of the unit cell array had both $x-$ and $y-$nodes grounded. Hence, voltages and hence fluxes at these nodes were set to zero, corresponding to the electronic analog of a zero-displacement (Dirichlet) boundary in the mechanical domain. The simple nature of the forcing (i.e. single amplitude/phase AC source) allows us to perform simulations directly in the frequency domain (small amplitude limit), in contrast to earlier experiments. Despite this, all the same realistic losses and tolerances were carried over from previous setups. The perspective taken here is of the OEC as a single lumped element between the source and the damping element of the right-most $x$-node. From this point of view, the relevant quantity is the amplification factor of the signal between the source and the right-most $x$-node, data of which is shown in Fig.~3b of the main text. Fig.~3c compares the power supplied through the source to the net elastic power supply/draw. Since passive elements do not perform any work, the power supply can be fully attributed to the internal active circuit elements.

The second amplification setup uses largely the same setup, but considers a $15\times 15$ lattice. The top, all boundary nodes are grounded, except the $x$-nodes on the left edge, which instead are subject to an identical single-phase AC signal. This corresponds to electrical imitation of a piston executing expansion-compression cycles. As with the minimal example, frequency domain analysis was carried out over a wide band of frequencies. The data for the large circuit is relatively high-dimensional - the double lattice contains more than 400 nodes, each of which has a complex amplitude (i.e. magnitude and phase offset) compared to the input signal, all representing highly nonlinear functions of forcing frequency. A summary of some features is provided in Figure~3d by exhibiting the RMS (in time) lattice site amplitudes at 4 distinguished frequencies: the lowest and the three strongest resonance peaks according to the mean (in space) oscillation amplitude in the system. The procedure for approximating resonance frequencies from ideal circuit theory is outlined in Sec.~B2 below.

\subsection{Ideal circuit theory of OECs}

In the following, we discuss the ideal circuit theory for OECs and derive the key analytic results presented in the main text.

\subsubsection{Apparent work extraction cycles}

In the section \textit{Apparent power extraction} of the main text, we compute elastic and dissipative work rates to illustrate the inherent chirality and non-reciprocity of the OEC and the potential for macroscopic work extraction from `microscopic' sources. Here, we present a more detailed computation to justify Eq.~(19) of the main text.

We will work on a lattice with odd side length $N=2M+1$, and choose lattice coordinates $\bsa \in \{-M,...,M\}^2$, such that the origin lies at the center of the lattice. We prove the odd case here, but the result holds for even sized lattices. First note that for a spatially constant flux gradient $\Phi$, it is straightforward to establish the spatio-temporally varying flux as 
\begin{equation}
    \phi_j^{\bsa} (t) = \alpha_i \Phi_{ij}(t) + c_j t + d_j,
\end{equation}
where $d_j$ are reference fluxes and $c_j$ are reference voltages, which correspond in the mechanical domain to specifying a reference frame for measurement. The voltage can then be expressed as
\begin{equation}\label{seq: site voltage}
    V_j^{\bsa} (t) = \alpha_i \dot\Phi_{ij}(t) +  c_j.
\end{equation}
Substituting homogeneous flux gradients of dilation-rotation cycles, Eq.~(18) in the main text, yields
\begin{equation}
    \begin{pmatrix}
        V_x^{\bsa}\\
        V_y^{\bsa}
    \end{pmatrix} = \Phi_0 \begin{pmatrix}
        \alpha_1 & \alpha_2
    \end{pmatrix} \begin{pmatrix}
        \omega \cos(\omega t) & \omega \sin(\omega t)\\
        -\omega\sin(\omega t) & \omega \cos(\omega t)
    \end{pmatrix} + \begin{pmatrix}
        c_1 \\ c_2
    \end{pmatrix},
\end{equation}
so Eq.~(17) from the main text can be expanded as
\begin{equation}
    \mathcal P_{\text{res}} = \frac{-\sigma \Phi_0^2 \omega^2}{TN^2} \oint dt \left( N^2 |\mathbf{c}|^2 +\sum_{\bsa} |\bsa|^2 \right).
\end{equation}
Note that for an analog shear-shear flux cycle one finds an equivalent expression. We have reduced the integrand to a time-independent expression, so it remains to evaluate the sum:
\begin{align}
    \sum_{\bsa} |\bsa|^2 &= \sum_{n=-M}^M \sum_{m=-M}^M n^2 + m^2 \nonumber\\
    &= 4N \sum_{n=1}^M n^2 \nonumber\\
    &= \frac{1}{6} N^2 (N^2 -1),
\end{align}
such that
\begin{equation}
    \mathcal P_{\text{res}} = -\sigma \Phi_0^2 \omega^2 \left( \frac{N^2-1}{6} + |\mathbf c|^2 \right).
\end{equation}
So, as claimed in the main text, the dissipation per lattice site can be expressed in terms of a constant $c_N$ which depends on the unspecified reference quantities and lattice size $N$, for which we attain the lower bound $c_N = (N^2-1)/6$ when we choose a frame co-moving with the center of the lattice.\\

Equation~(16) of the main text, describing the power output $\mathcal{P}_{\text{ind}}$ of inductive components, serves to highlight the role of non-mutual currents in a power extraction. To compute a simplified expression for $\mathcal{P}_{\text{ind}}$ for the case of cyclic excitations it will be more convenient to use previously obtained facts about the circuit, to avoid dealing explicitly with the large (though sparse) admittance matrix $\boldsymbol{\Lambda}$. Note that decomposition into Eqs.~(16), (17) uses linearity of power with respect to current for given voltages. We can further split the power using the observation that currents cancel for homogeneous flux gradients at bulk sites, and compute only the power output from one boundary, obtaining the full boundary by superposition. In particular, for the positive $x$-boundary the `missing' boundary node fails to supply currents $\mathcal I_{xx} = k_p \Phi_{xx}$ to the $x$-node and $\mathcal I_{xy} = - k_a \Phi_{xx}$ to the $y$-node (see Eq.~(10), main text). Using Eq.~(\ref{seq: site voltage}) and ignoring the reference flux that must drop out by symmetry, we find
\begin{align}
    \mathcal P_{\text{ind}} &= \frac{4}{TN^2} \oint dt \sum_{m=-M}^M \begin{pmatrix}
        M \\ m
    \end{pmatrix}^T \mathbf{\dot\Phi} \begin{pmatrix}
        -k_p \\  k_a
    \end{pmatrix} \Phi_{11}\nonumber\\
    &= \frac{4}{TN^2} \Phi_0^2 \omega \oint dt \sin(\omega t) \begin{pmatrix}
        N(N-1)/2 \\ 0 
    \end{pmatrix}^T \begin{pmatrix}
        -k_p \cos(\omega t) + k_a \sin(\omega t)\\
        k_p \sin(\omega t) + k_a \cos(\omega t)
    \end{pmatrix}\nonumber\\
    &= \Phi_0^2 \omega k_a (1-N^{-1}).
\end{align}
We thus recover the frequency-dependent power extracted from the circuit as 
\begin{equation}\label{seq: P omega}
\mathcal{P}(\omega)=P_{\text{res}}(\omega)+P_{\text{ind}}(\omega)= \Phi_0^2\left[(1-N^{-1})k_{\text{a}}\omega - c_{N}\sigma\omega^2\right],
\end{equation}
whose maximum defines the frequency of maximal gain $\omega^* = 3 k_a /[\sigma N (N+1)]$.

Finally, we consider the adiabatic limit $\omega\rightarrow0$ and compute the extracted work done per cycle, which yields
\begin{equation} \label{seq: adiabatic}       \mathcal{W}=\lim_{\omega\rightarrow0^{\pm}}T\mathcal{P}=\pm (1-N^{-1})2\pi\phi_0^2k_{\text{a}}.
\end{equation}
An OEC, as proposed here, therefore realizes the theoretical concept of an active engine~\hbox{\cite{etien2021b,Sousl2021}} that hypothetically can achieve a perfect energy conversion efficiency -- at the expense of ignoring internal active circuit elements that facilitate this apparent gain. Additionally, Eq.~(\ref{seq: adiabatic}) recapitulates the notion that the amount of extracted work is proportional to the area enclosed by the cycle in strain space~\cite{Scheibner:2020gm,tan2022}, in our case the area $2\pi\phi_0^2$ enclosed in the space of flux gradient components.

\subsubsection{Amplification and active resonance}\label{ssec: amp theory}

We define internal nodes $(\bsa,i)$ as those that are not at the circuit boundary and do not see a prescribed flux due to an external input signal. For these nodes, Eq. (13) of the main text reads as an evolution equation,
\begin{equation}
    \sigma\dot\phi_i^{\bsa} = - \tilde\Lambda_{ij}^{\bsa \bsb} \phi_j^{\bsb} - f_i^{\bsa},
    \label{eq: evolution eqn}
\end{equation}
where $\tilde\Lambda_{ij}^{\bsa \bsb}$ is the modified inductance matrix (Eq.~(14), main text) and $f_i^{\bsa}$ are currents arriving at $(\bsa,i)$ due to interactions with boundaries or forced nodes. Neglecting transients, this can be re-expressed in the frequency domain,
\begin{equation}
    \left( i\omega\sigma \delta_{ij}^{\bsa \bsb} +  \tilde \Lambda_{ij}^{\bsa \bsb} \right) \hat\phi_j^{\bsb} = - \hat f_i^{\bsa}.
    \label{eq: evolution f space}
\end{equation}
Equation~(\ref{eq: evolution f space}) provides an implicit (nonlinear) relationship between the response amplitudes $\hat\phi_i^{\bsa}$ and the forcing frequency, which we consider as the response function. The solutions to this system of equations form the basis for theoretical predictions of amplification and resonances, which we have compared to realistic circuit simulations in the main text (see Fig.~3). 

In the following, we explain how to obtain closed form expressions for resonant frequencies that were compared against realistic circuit simulations resonances in Fig.~3d (dashed lines) of the main text. For the square lattice with nearest-neighbor interactions, the modified coupling matrix $\boldsymbol{\tilde \Lambda}$ can be related to graph Laplacians of the underlying spring networks and, in particular, one can show that eigenvectors of the evolution operator are discrete sines and cosines (up to a weighting between $x$- and $y$-components).

Consider first the two graphs corresponding to the connectivity structure of interior $x$ and $y$-nodes respectively, and let their graph Laplacians (modified for Dirichlet boundary conditions) be denoted $L_x$ and $L_y$. We may make this choice of boundary data since the discrete problem can have boundary forces built in manually into the $\hat f$ term in Eq.~(\ref{eq: evolution f space}). The key observations are that the Laplacians serve to describe the flux-current relations (up to effective stiffnesses), as well as that the two graphs are Cartesian graph products of the $N$-node line graph and the edgeless graph of size $N$. Letting $L_0$ be the Laplacian for the $N$-node line graph with Dirichlet boundary, and letting $I$ denote the identity of appropriate size, then graph product identities~\cite{Merris1994} allow us to express $L_x$ and $L_y$ as Kronecker products,
\begin{align}
    L_x &= L_0 \otimes I, \\
    L_y &= I \otimes L_0.
\end{align}
In particular, letting $j\in\{1,...,N\}$ and $(\lambda_j,\mathbf v_j )$ be eigenvalue-vector pairs for $L_0$, then $L_x$ and $L_y$ share eigenvectors $\mathbf u_{jk} = \mathbf v_j \otimes \mathbf v_k$, with eigenvalues $\lambda_j$ and $\lambda_k$ respectively. There exists a closed form for $\lambda_j$~\cite{Chung2000},
\begin{equation}
    \lambda_j = 4 \sin^2\left( \frac{\pi j}{2(N-1)}\right),
\end{equation}
where the grid is $N \times N$ and the first and last row and column have the Dirichlet boundary condition imposed, i.e. $(N-2)^2$ interior nodes. Making explicit the connection between graph Laplacians and currents, one can assemble a block matrix $L$, a reordering of $\boldsymbol{\tilde \Lambda}$:
\begin{equation}
    L = \begin{pmatrix}
        k_p L_x & -k_a L_y \\
        k_a L_x & k_p L_y
    \end{pmatrix}.
\end{equation}
Using $\mathbf u$ and $\mathbf f$ to denote the appropriate vectorization of components $\hat \phi_j^{\bsb}$ and $\hat f_j^{\bsb}$, and again $I$ for the appropriately sized identity, then Eq.~(\ref{eq: evolution eqn}) can be rewritten,
\begin{equation}
    (i\omega I + L)\mathbf u = \mathbf f.
\end{equation}
The power of this approach is that we have reduced the problem to a system of linear equations, for which we have information about eigenvalue/vector pairs of blocks. It is precisely this formalism which allows to symbolically (via MATLAB Symbolic Toolbox) obtain the ideal response in Fig.~3b of the main text.

Note also now that we need only find eigenvectors for $L$ to characterize the spectrum of the evolution operator, and we already have eigenvectors of its constituents $L_x$ and $L_y$. We consider a concatenation of scaled copies of such eigenvectors, with scale factor $\alpha\in \mathbb C$,

\begin{equation}
    L \begin{pmatrix}
        \mathbf v_j \otimes \mathbf v_k \\
        \alpha \mathbf v_j \otimes \mathbf v_k
    \end{pmatrix} = \begin{pmatrix}
        (k_p \lambda_j - \alpha k_a \lambda_k) \mathbf v_j \otimes \mathbf v_k \\
        (k_a \lambda_j + \alpha k_p \lambda_k) \mathbf v_j \otimes \mathbf v_k 
    \end{pmatrix},
\end{equation}
and hence we find that $\alpha$ solves a quadratic equation,
\begin{equation}
    \alpha^2 k_a \lambda_k + \alpha k_p (\lambda_k - \lambda_j) + k_a \lambda_j = 0.
\end{equation}

For example, the $j$\textsuperscript{th} diagonal mode of the evolution operator has $\alpha = \pm i$, then then the corresponding eigenvalue of the evolution operator $(i\omega I + L)$ is
\begin{equation}
    \xi_j = i\omega\sigma + 4 \sin^2 \left( \frac{\pi j}{2(N-1)}\right) (k_{\text{p}} \pm ik_{\text{a}}),
    \label{eq: evolution evals}
\end{equation}
which bears a striking resemblance to the continuum dispersion relation Eq.~(\ref{seq:disp rel}). In general, i.e. for non-diagonal modes, the form of the eigenvalues is more complicated and is omitted here for brevity.\\

One can even restrict this framework to obtain the passive response ($k_{\text{a}}=0$) to see that the response is symmetric around $\omega=0$, as we would expect from a passive, reciprocal theory. In fact, decoupling of $x$/$y$ fluxes leads the system to respond only as rows of $N-1$ springs in series, which behaves as a single spring with some effective stiffness. This is the basis for the scaling of the passive curve in Fig.~3b, for which the theoretical and experimental data were in such good agreement that they could not visually distinguished in the figure, and hence only one was included. \\

Equation (\ref{eq: evolution evals}) forms the basis for a qualitative understanding of the large amplitude regime, since large amplification can only occur when the evolution operator is close to singular. The resonant frequency is determined by minimizing the imaginary part $\text{Im}(\xi_j) = \omega\sigma \pm  \lambda_j k_{\text{a}}$, and correspondingly the bandwidth is on the order of $\lambda_j k_p/\sigma$, from comparison to the real part. In this way, we are able to rationalize the experimental observations in the $15\times 15$ OEC, by identifying the peaks as these resonances (Fig.~3d), matching first three resonance peaks to good accuracy. Note that the symmetry of the forcing restricts to odd wavenumber in the vertical direction, i.e. because only these components of the forcing are excited. A less structured forcing term may excite other resonances in this range. 


%